\documentclass[12pt, a4paper]{article}
\pdfoutput=1
\usepackage[dvipdfmx]{graphicx}
\usepackage{amssymb}
\usepackage{amsmath}
\usepackage{bm}
\usepackage{comment}
\usepackage{color}
\usepackage{cite}
\usepackage{slashed}
\usepackage{subfigure}
\usepackage{epstopdf}            
\usepackage{epsfig}
\renewcommand{\thefootnote}{\fnsymbol{footnote}}
\newcommand{\beq}{\begin{equation}}
\newcommand{\eeq}{\end{equation}}
\newcommand{\pa}{\partial}

\newcommand{\Slash}[1]{{\ooalign{\hfil/\hfil\crcr$#1$}}}

\newcommand{\Tr}{\text{Tr}}

\usepackage[colorlinks=true, linkcolor=black, citecolor=black,
urlcolor=black]{hyperref}

\begin{document}

\begin{titlepage}

\begin{center}

{\Large
{\bf 
WIMP dark matter in the parity solution to the strong CP problem\\}
}
\vskip 2cm

Junichiro Kawamura$^{1,2,}$\footnote{kawamura.14@osu.edu},
Shohei Okawa$^{3,4,}$\footnote{okawa@uvic.ca},
Yuji Omura$^{5,}$\footnote{yomura@phys.kindai.ac.jp},
Yong Tang$^{6,}$\footnote{ytang@hep-th.phys.s.u-tokyo.ac.jp}

\vskip 0.5cm

{\it $^1$Department of Physics,
The Ohio State University, 
%191 W. Woodruff Ave, 
Columbus, OH 43210, USA}\\[3pt]

{\it $^2$Department of Physics, 
Keio University, Yokohama 223-8522,  Japan} \\[3pt]

{\it $^3$ Physik Department T30d, Technische Universit\"{a}t M\"{u}nchen, \\
James-Franck-Stra$\beta$e, 85748 Garching, Germany}

{\it $^4$ Department of Physics and Astronomy, University of Victoria, \\
Victoria, BC V8P 5C2, Canada}

{\it $^5$
Department of Physics, Kindai University, Higashi-Osaka, Osaka 577-8502, Japan}\\[3pt]

{\it $^6$Department of Physics, University of Tokyo, 
Tokyo 113-0033, Japan} \\[3pt]

\vskip 1.5cm

\begin{abstract}
We extend the Standard Model (SM) with parity symmetry, motivated by the strong CP problem and dark matter. 
In our model, parity symmetry is conserved at high energy by introducing a mirror sector with the extra gauge symmetry, $ SU(2)_R \times U(1)_R$. The charges of $SU(2)_R \times U(1)_R$ are assigned to
the mirror fields in the same way as in the SM, but the chiralities of the mirror fermions are opposite to respect the parity symmetry. The strong CP problem is resolved, since the mirror quarks are also charged under the $SU(3)_c$ in the SM. In the minimal setup, the mirror gauge symmetry leads to stable colored particles   
which would be inconsistent with the observed data,
so that we introduce two scalars in order to deplete the stable colored particles. 
Interestingly, one of the scalars becomes stable because of the gauge symmetry and therefore can be a good dark matter candidate. We especially study the phenomenology relevant to the dark matter, 
i.e. thermal relic density, direct and indirect searches for the dark matter. The bounds from
the LHC experiment and the Landau pole are also taken into account.
As a result, we find that a limited region is viable: the mirror up quark mass is around [600\,GeV, 3\,TeV] and the relative mass difference between the dark matter and the mirror up quark or electron is about ${\cal O}$(1-10\,\%).
We also discuss the neutrino sector and show that the right-handed neutrinos in the mirror sector can increase the effective number of neutrinos or dark radiation by $0.14$. 
\end{abstract}

\end{center}
\end{titlepage}

\renewcommand{\thefootnote}{\#\arabic{footnote}}
\setcounter{footnote}{0}

\section{Introduction}

The Standard Model (SM) is very successful in explaining enormous results at terrestrial laboratories. In particular, the LHC discovered the Higgs boson that was the last piece of the SM. In the meantime, various cosmological and astrophysical observations indicate that the SM has to be extended in order to account for dark matter (DM), neutrino oscillation, baryon asymmetry and so on. There are also several theoretical issues in the intrinsic structure of the SM. For example, the SM cannot explain why the mass of Higgs boson is extremely small compared with the Planck scale, namely the gauge hierarchy problem.   

Another issue of the SM is the so-called strong CP problem. 
The SM gives no reason why the coefficient for the $\theta$-term,   
 \begin{align}
  \theta\dfrac{g^2_s}{32\pi^2}F^a_{\mu\nu}\tilde{F}^{a\mu\nu}, 
\end{align}
is so tiny to be consistent with the experimental limit 
$|\theta|\lesssim 10^{-10}$~\cite{Baker:2006ts}. 
The SM consists of quarks, leptons and Higgs field charged under the 
$SU(3)_c \times SU(2)_L \times U(1)_Y$ gauge symmetry. 
The parity symmetry is explicitly broken, 
since right-handed fermionic fields are $SU(2)_L$ singlets 
while left-handed ones are doublets. 
In addition, CP symmetry is also broken by the complex Yukawa matrices. 
Thus the CP-violating $\theta$-term is also legitimately allowed, 
and $\theta$ is expected to be $\mathcal{O}(1)$. 

This problem may be a good clue to consider new physics beyond the SM.  
There have been many attempts to solve this problem 
by introducing the Peccei-Quinn symmetry and axion~\cite{Peccei:1977hh, Peccei:1977ur, Weinberg:1977ma, Wilczek:1977pj}, left-right symmetry~\cite{Babu:1989rb,Barr:1991qx,Gu:2012in,Abbas:2017vle,Abbas:2017hzw}. 
These extensions have been discussed with applications to neutrino physics~\cite{Gu:2011yx}, the baryon asymmetry~\cite{Gu:2013gic,Gu:2017mkm}, the LHC physics~\cite{DAgnolo:2015uqq,Alvarado:2018acf}, grand unification~\cite{Hall:2018let}, flavor physics~\cite{Ema:2016ops,Ema:2018abj,Calibbi:2016hwq,Alanne:2018fns} and DM physics~\cite{Alves:2016bib,Dev:2016xcp}. 

In this paper, we propose a model that the SM has its mirror sector, 
so that a parity symmetry is respected at high energy scale. 
The SM gauge symmetry $SU(3)_c \times SU(2)_L \times U(1)_Y$ 
is extended to $SU(3)_c \times SU(2)_L \times U(1)_L \times SU(2)_R \times U(1)_R$, 
and the mirror fermions charged under $SU(3)_c  \times SU(2)_R \times U(1)_R$ 
are introduced in this model.   
This parity symmetry forbids the $\theta$-term until it is spontaneously broken at some low energy scale. 
A nonzero $\theta$-term is then induced through the renormalization group (RG) running, 
but its value is still controlled and kept tiny even at low energy scale.  

Scalar fields are introduced in addition to the above minimal parity-symmetric model, 
otherwise one of the mirror quarks or leptons would be stable as a result of the gauge symmetry, 
since there is no portal coupling between the SM and the mirror sector \cite{Perl:2001xi}. 
The scalar fields can mediate decays of the mirror fermions and deplete them in the early universe. 
Interestingly, the lightest scalar can be neutral under the SM gauge symmetry 
and can also be stable due to the remnant of the extended gauge symmetry. 
Therefore this scalar field becomes a good DM candidate.  

One of our motivations in this paper is to study the thermal relic density of the scalar field 
and constraints from the DM searches. 
The DM physics is closely related to the mirror fermion masses 
and the Yukawa couplings among the scalar fields and the mirror fermions,  
since the mirror fermions act as mediators in the scattering of the DM with SM particles. 
A remarkable feature of this mirror model is 
that the ratios among masses of the mirror fermions are the same as in the SM 
above the parity breaking scale. 
All the mirror fermions will get upper bounds on their masses, 
once any of their masses are constrained by DM observations. 
In addition, the parity breaking scale could have an upper bound 
in order to keep the perturbativity up to this scale. 
In particular, we show that the mirror up quark should be lighter than a few TeV in Sec. \ref{sec;phenomenology}. 
This result indicates that the parity breaking scale should be below $4\times10^8$\ GeV. 
The figures in Sec. \ref{sec;phenomenology} also suggest that the mass splitting of DM 
and a mediator fermion has to be ${\cal O}$(1-10\,\%) to account 
for the observed density, while the DM-nucleus scattering cross section is smaller than the limit
from the DM direct direction experiment.

This paper is organized as follows. 
In Sec.~\ref{sec;setup} we introduce our model with the parity symmetry. 
In Sec.~\ref{sec;cp} we show how this model can solve the strong CP problem. 
Then in Sec.~\ref{sec;phenomenology} we discuss phenomenological aspects of this model:
the DM physics, the flavor physics and the LHC physics. 
In Sec.~\ref{sec;neutrino} we study the neutrino sector and discuss possible impacts on cosmological observables. 
Finally, we summarized the results in Sec.~\ref{sec;summary}.
In Appendix, we show the relevant RG equations and investigate the effects of a Higgs portal coupling on the DM physics.

\section{The model with the parity symmetry}
\label{sec;setup}

In this section, 
we shall construct an extended model with $SU(3)_c \times SU(2)_L \times U(1)_L \times SU(2)_R \times U(1)_R$ which respects the parity symmetry. 
The parity symmetry forbids the $\theta$-term. 
We shall also establish our conventions and notations here. 
In general, the parity transformation of a Dirac fermion field, $q_i$ ($i=1, \dots, N_F$), 
is defined as
 \begin{align}
\label{eq-parity}
P \, q^i(t, x) \, P =  \gamma_0 \, q^i (t,-x),
 \end{align}
using the parity operator $P$, that satisfies $P^2=1$~\footnote{We have adopted the convention $P=P^{-1}$.}.
The Lagrangian density $\mathcal{L}_q$ for $q^i$ and a scalar $\Phi$ is given by
\beq
{\cal L}_q= \left(\overline{q^i} \, \gamma^\mu D_\mu \, q^i - y^{ij} \,  \overline{q^i} \, \Phi \, q^j +h.c.\right) + \frac{1}{2} \, \Tr_4 \left[ \left ( D^\Phi_\mu \Phi \right)^\dagger D_{\Phi}^{ \mu} \Phi \right ]  -V (\Phi) ,
\eeq
which can be invariant under the parity transformation~Eq.(\ref{eq-parity}) 
depending on the covariant derivatives, $D_\mu$ and $D^\Phi_\mu$ 
and the scalar field $\Phi$. 
Let us consider a parity symmetric gauge group $G_L\times G_R$, 
where $G_{L}$ can be a product group like $SU(2)_L\times U(1)_L$ and $G_R$ as well. 
The covariant derivatives are given by,  
\begin{eqnarray}
D_\mu = \partial_\mu + i g_I {\cal Q}^I_q  A^I_\mu, 
\quad
D^\Phi_\mu = \partial_\mu + i g_I {\cal Q}^I_\Phi A^I_\mu,
\end{eqnarray}
where $A^I_\mu$ is the gauge field for the gauge groups $I=G_L, G_R$, 
$g_I$ denotes the gauge coupling constant and ${\cal Q}^I_{q,\Phi}$ is a representation matrix for a non-Abelian group or a charge for $U(1)$.

Here, the gauge field and scalar field are defined as
\begin{eqnarray}
A^{I \, \mu}= \begin{pmatrix} A_R^{I \, \mu} & 0 \\ 0 &  A_L^{I \, \mu} \end{pmatrix}, ~\Phi&=& \begin{pmatrix}  H_R & 0 \\ 0 &   H_L  \end{pmatrix}. 
\end{eqnarray}
The gauge field $A^I_\mu$ and the scalar field $\Phi$ are the $4 \times 4$ matrices 
that do not commute with $\gamma_\mu$ 
and each of the elements is linear to a $2\times 2$ unit matrix.
Note that the fermion $q_i$ is decomposed 
into the right-handed and the left-handed components, i.e. 
$q^i= (q^i_R, \, q^i_L)^T$, in this description. 
$A_R^{I \, \mu} $ and $A_L^{I \, \mu}$ correspond to the gauge fields 
for the gauge symmetry that act on $q^i_R$ and $q^i_L$, respectively.
The scalar field $\Phi$ and the gauge field $A^I_\mu$ are transformed under the parity as
\begin{eqnarray}
P \, \Phi(t, x) \, P&=&\gamma_0 \, \Phi(t, -x) \, \gamma_0=   \begin{pmatrix}  H_L (t, -x) &0  \\  0&  H_R (t, -x)  \end{pmatrix}, \label{trans1}  \\
P \, A^I_\mu(t, x) \, P&=&\gamma_0 \, A^{I \, \mu}(t, -x) \, \gamma_0= \begin{pmatrix} A_L^{I \, \mu}  (t, -x) & 0 \\ 0 &  A_R^{I \, \mu}  (t, -x)   \end{pmatrix}. \label{trans2}
\end{eqnarray}
Thus, the parity transformation leads the following exchange:
\beq
A_{L \, \mu}(t, x) \leftrightarrow A^{\mu}_R (t,-x), \, H_L (t,x) \leftrightarrow H_R (t,-x).
\eeq
The Lagrangian ${\cal L}_q$ becomes invariant under the parity transformation as far as the $\gamma_0$ dependence does not show up explicitly in a scalar potential $V (\Phi)$.
The gauge interaction should also respect the parity symmetry:
\beq
{\cal L}_g= -\frac{1}{4 } \Tr_4 \left ( F^{I \, \mu \nu} F^I_{ \mu \nu}  \right ) ,
\eeq
where $F^I_{\mu \nu}$ is the field strength composed by $A^I_\mu$. One can immediately see that the $\theta$-term is not allowed by the parity symmetry. 
Note that there are two chiral gauge symmetries described by $A^{I \, \mu}_L$ and $A^{I \, \mu}_R$ that have the same gauge coupling. In addition, we can find the gauge kinetic term in the $U(1)$ gauge symmetry case.

%%%%%%%%%%%%%%%%%%%%%%%%%%%%%%%%%%%%%%%%%%%%%%%%%%%%%%%%%%%
\begin{table}[t]
\begin{center}
\begin{tabular}{ccccccc}
\hline
\hline
Fields & spin   & ~~$SU(3)_c$~~ & ~~$SU(2)_L$~~  & ~~$SU(2)_R$~~  & ~~$U(1)_R$  & ~~$U(1)_L$      \\ \hline  
  $Q^{i }_{L}$  & $1/2$ & ${\bf 3}$         &${\bf2}$    &   ${\bf1}$ &      $0$   &       $1/6$       \\  
   $u^{i }_{ R}$  & $1/2$&  ${\bf 3}$        &${\bf1}$&   ${\bf1}$   &       $0$    &         $2/3$        \\ 
     $d^{i }_{ R}$ & $1/2$ &  ${\bf 3}$        &${\bf1}$ &   ${\bf1}$   &       $0$   &         $-1/3$       \\  \hline 
     $l^{i }_{L}$ & $1/2$ & ${\bf 1}$         &${\bf2}$ &   ${\bf1}$  &       $0$    &       $-1/2$    \\  
   $e^{i }_{ R}$& $1/2$  &  ${\bf 1}$        &${\bf1}$ &   ${\bf1}$   &       $0$   &         $-1$        \\  \hline
   $H_L$& $0$  & ${\bf1}$         &${\bf 2}$ &   ${\bf1}$  &       $0$    &            $1/2$             \\ \hline \hline
\end{tabular}
\end{center}
\caption{Matter content in the SM sector. $i$ denotes the flavors: $i=1, \, 2, \, 3$. }
 \label{table1}
\end{table} 
%%%%%%%%%%%%%%%%%%%%%%%%%%%%%%%%%%%%%%%%%%%

Based on this generic argument, we extend the SM to the parity conserving model.
In the SM, the gauge symmetry is $G_{SM}=SU(3)_c \times SU(2)_L \times U(1)_Y$. Now,
we extend the gauge symmetry as 
\beq
G_{SM} \to SU(3)_L \times SU(3)_R \times SU(2)_L \times SU(2)_R \times U(1)_L \times U(1)_R .
\eeq
Since the SM is vector-like under $SU(3)_c$ and the parity symmetry is respected in this interaction, 
we only consider the diagonal direction of $SU(3)_L \times SU(3)_R$, 
that is identified to $SU(3)_c$. 
Based on the argument above,
the parity transformation leads the exchange of the symmetry:
\beq
SU(2)_L \times U(1)_L \leftrightarrow SU(2)_R \times U(1)_R.
\eeq
This extension has been proposed to solve the strong CP problem \cite{Barr:1991qx} which however did not have a DM candidate.

The matter content in the SM sector is summarized in Table \ref{table1}. 
$Q^{i }_{L}$,  $u^{i }_{ R}$, and $d^{i }_{ R}$ ($i=1,\,2,\,3$) correspond
to the SM quarks charged under $SU(3)_c \times SU(2)_L \times U(1)_L$. 
The fields, $l^{i }_{L}$ and $e^{i }_{ R}$, denote the leptons. 
$H_L$ is the scalar field that causes the electroweak (EW) symmetry breaking.
The relevant Yukawa interactions are written as
\begin{equation}
  {\cal L}_L=- Y^{ij}_d  \overline{Q^i_L} H_L d_R^j  -  Y^{ij}_u \overline{Q^i_L} \widetilde H_L u_R^j 
  - Y^{ij}_e \overline{l^i_L} H_L  e_R^j +h.c.,
\end{equation}
where $ \widetilde H_L$ is defined as $ \widetilde H_L = i \sigma_2 H_L^*$.  
$\sigma_2$ is the Pauli matrix.

We introduce a mirror sector to respect the parity symmetry as follows. 
The matter content of the mirror sector is summarized in Table \ref{table2}.
$Q^{\prime \, i }_{L}$,  $u^{\prime \, i }_{ R}$, and $d^{\prime \, i }_{ R}$ ($i=1,\,2,\,3$) are the mirror quarks charged under $SU(3)_c \times SU(2)_R \times U(1)_R$.
The fields, $l^{\prime \, i }_{L}$ and $e^{\prime \, i }_{ R}$, denote the mirror leptons.
$H_R$ is a scalar charged under $SU(2)_R$ but not under $SU(2)_L$.
The vacuum expectation value (VEV) plays a role in making the mass hierarchy between the
SM and mirror sectors. The detail will be shown below. 
%%%%%%%%%%%%%%%%%%%%%%%%%%%%%%%%
%%%%%%%%%%%%%%%%%%%%%%%%%%%%%%%%%%%%%%%%%%%%%%%%%%%%%%%%%%%
\begin{table}[t]
\begin{center}
\begin{tabular}{ccccccc}
\hline
\hline
Fields & spin   & ~~$SU(3)_c$~~ & ~~$SU(2)_L$~~& ~~$SU(2)_R$~~  & ~~$U(1)_R$  & ~~$U(1)_L$      \\ \hline  
  $Q^{ \prime \, i}_{R}$  & $1/2$ & ${\bf 3}$   &${\bf1}$       &${\bf2}$      &       $1/6$ &       $0$      \\  
   $u^{\prime \, i}_{ L}$  & $1/2$&  ${\bf 3}$   &${\bf1}$      &${\bf1}$      &         $2/3$       &       $0$  \\ 
     $d^{\prime \, i}_{ L}$ & $1/2$ &  ${\bf 3}$   &${\bf1}$      &${\bf1}$      &         $-1/3$   &       $0$      \\  \hline 
     $l^{\prime \, i}_{R}$ & $1/2$ & ${\bf 1}$   &${\bf1}$       &${\bf2}$      &       $-1/2$   &       $0$   \\  
   $e^{\prime \, i}_{ L}$& $1/2$  &  ${\bf 1}$  &${\bf1}$       &${\bf1}$      &         $-1$    &       $0$     \\  \hline
   $H_R$& $0$  & ${\bf1}$   &${\bf1}$       &${\bf 2}$      &            $1/2$     &       $0$     \\ \hline \hline
\end{tabular}
\end{center}
\caption{Matter content in the mirror sector. $i$ denotes the flavors: $i=1, \, 2, \, 3$. }
 \label{table2}
\end{table} 
%%%%%%%%%%%%%%%%%%%%%%%%%%%%%%%%%%%%%%%%%%%

The Yukawa couplings among the mirror fields are written down as follows:
\begin{equation}
  {\cal L}_R=- Y^{ij}_d  \overline{Q^{\prime \, i}_R} H_R d^{\prime \, j}_L  -  Y^{ij}_u \overline{Q^{\prime \, i}_R} \widetilde H_R u_L^{\prime \, j} 
  - Y^{ij}_e \overline{l^{\prime \, i}_R} H_R  e_L^{\prime \, j} +h.c..
\end{equation}
Note that the Yukawa couplings are defined to respect the parity symmetry, that corresponds to the following exchange:
\begin{eqnarray}
\label{eq;LRsymmetry}
Q^i_L (t,x) \leftrightarrow Q^{\prime \, i}_R (t,-x), \, && u^i_R (t,x) \leftrightarrow u^{\prime \, i}_L (t,-x), \, d^i_R (t,x) \leftrightarrow d^{\prime \, i}_L (t,-x),  \nonumber \\
l^i_L (t,x) \leftrightarrow l^{\prime \, i}_R (t,-x), \, && e^i_R (t,x)\leftrightarrow e^{\prime \, i}_L(t,-x), \, H_R (t,x)\leftrightarrow H_L(t,-x). 
\end{eqnarray}

The structure of the mirror sector is the same as the one of the SM sector, because of the parity symmetry.
Then, we expect that some stable particles appear in the mirror sector in the same way 
as the proton and electron in the SM.
Those stable particles are, however, 
strongly constrained by cosmological observations 
and searches for stable extra charged particles. 
Below, we discuss the stability of the mirror particles and
investigate a possibility that some neutral particles become cold DM candidates.
After that, we propose one extension to avoid these stable charged particles.

\subsection{Stability of the extra particles and dark matter candidate}
In our model, $SU(2)_L \times U(1)_L \times SU(2)_R \times U(1)_R$ breaks down to the EW symmetry.
The VEV of $H_L$ breaks down the EW symmetry 
to the electromagnetic (EM) symmetry, $U(1)_{em}$.
Let us consider the case where the EM charge of the field $q$, 
${\cal Q}^q_{em}$, is given by
\beq
\label{bad scenario}
{\cal Q}^q_{em} = \tau^q_L+{\cal Q}^q_L, 
\eeq
where $\tau^q_L$ is the isospin given by the third component of $SU(2)_L$ and ${\cal Q}^q_L$ is
the $U(1)_L$ charge. In this case, the mirror particles are not charged under $U(1)_{em}$.
In our model, the non-vanishing VEV of $H_R$ breaks 
$SU(2)_R \times U(1)_R$ down to the mirror $U(1)_{em}^m$, 
which is orthogonal to the EM symmetry in the SM.

We find that the mirror quarks cannot decay to the SM quarks in this scenario.
For convenience, let us define the subgroup of $U(1)_{em}$:
$U(1)_{em} \supset Z^{em}_3$.
In the SM, the up-type quarks $u_i$ and down-type quarks $d_i$
are charged under $SU(3)_c \times Z^{em}_3$ as follows:
\beq
u_i \, : \, ( {\bf 3}, \omega), \quad d_i \, : \, ( {\bf 3}, \omega),
\eeq
where $\omega^3=1$ is satisfied in our notation.
We note that the other $SU(3)_c$-singlet fields in the SM are not charged under the $Z^{em}_3$ symmetry. 
Any $SU(3)_c$-singlet composite operators that consist only of the SM fields 
are not charged under the $Z^{em}_3$
~\footnote{This can be easily understood by using Young tableau.
One $\Box$ carries $\omega$ charge in the SM quark sector.
$SU(3)_c$ invariance requires $3 \times N$ $\Box$ ($N=1,\,2, \dots$), so that
the $SU(3)_c$-singlet operators are $Z^{em}_3$-singlet in the SM.}.
The mirror quarks and leptons are, on the other hand, not charged under the $Z^{em}_3$ in this scenario,
since they are $U(1)_{em}$-singlet.
The mirror quarks $u^\prime_i$ ($d^\prime_i$) cannot decay unless 
there exists $SU(3)_c$-singlet operator 
which contains only one $u^\prime_i$ ($d^\prime_i$). 
Such a $SU(3)_c$-singlet operator involving one mirror quark 
is, however, always charged under the $Z^{em}_3$ symmetry.
Thus, the lightest mirror quark becomes stable  
unless extra $Z^{em}_3$-charged fields are introduced. 
When such an extra $Z^{em}_3$-charged field is introduced, it becomes stable 
due to the $U(1)_{em}$ and $Z^{em}_3$ symmetry.

The remnant symmetry of $U(1)_R$ also makes some particles stable.    
If only $ \langle H_R \rangle$ breaks $SU(2)_R \times U(1)_R$,
the $U(1)^R_{em}$ symmetry remains in the same manner 
as the EW symmetry breaking.
The gauge symmetry forbids the lightest mirror quark and the mirror electron to decay.
Even if we introduce some scalar fields charged under 
$SU(2)_R$ and/or $U(1)_R$ gauge symmetry
to break the $U(1)^R_{em}$ symmetry spontaneously, 
the remnant symmetry from $U(1)^R_{em}$
would guarantee the stability of the $U(1)^R_{em}$-charged particles.

Such stable mirror particles may lead to unfavorable consequences. 
At the QCD (de)confinement transition in the early universe, 
the lightest mirror quark would form stable exotic hadrons 
together with the SM light quarks, such as $q^\prime \bar{q}$ and $q^\prime q q$. 
Since these hadrons are fractionally charged and 
scatter with visible matter via the strong or EW interactions, 
the cosmological abundance is strongly constrained. 
For stable colored particles much heavier than the confinement scale, 
the abundance of the exotic hadrons has been estimated in 
the literature~\cite{Kang:2006yd,Jacoby:2007nw,Geller:2018biy}, 
taking non-perturbative effects at or below the QCD scale into account, as 
\begin{equation}
\Omega_{\rm exotics} h^2 
\sim \sqrt{ \frac{\Lambda_{\rm QCD}}{m} } \left( \frac{m}{30\,{\rm TeV}} \right)^2 ,
\end{equation}
where $m$ and $\Lambda_{\rm QCD}$ denote the colored particle mass and 
the QCD confinement scale, respectively. 
This gives a small value of ${\cal O}(10^{-4})$ for $m={\cal O}$(TeV), 
while the direct searches for strongly interacting particles and fractionally charged particles 
will put severe constraints on their flux at the Earth surface~\cite{Perl:2001xi}. 
Note that the precise prediction for the cosmological abundance and the experimental bounds 
require a full knowledge of the non-perturbative QCD. 
Thus, further careful studies are needed to conclude the viability of this scenario, 
and it will be pursued elsewhere. 

In addition, the mirror electron is also stable in this case. 
The thermal abundance set by $e^\prime \bar{e}^\prime \to \gamma' \gamma'$ is estimated as 
$\Omega_{e'} h^2 \simeq 0.1 ( m_{e'} / 100\,{\rm GeV} )^2$. 
The mirror electron mass is correlated with the mirror up quark mass,  
since their masses are given by the Yukawa coupling constants  
that are fixed by the SM Yukawa coupling constants at the parity breaking scale. 
The LHC limit on the mirror up quark comes from the search for the so-called R-hadron 
which is a composite state involving supersymmetric particles.   
The lower limit on a top squarks mass is about 890 GeV~\cite{Aaboud:2016uth} 
from the search for the R-hadrons coming from top squark pair production. 
The limit on the mirror up quark in this model is estimated as 1080 GeV 
by assuming the pair production cross section of the mirror up quark is four times as that of top squarks. 
This means that the mirror electron should be heavier than 250 GeV, 
and then the relic would be too abundant.
Besides, if there is a gauge kinetic mixing between $U(1)_L$ and $U(1)_R$, 
the mirror electron can be millicharged. 
The stable millicharged particle can affect the CMB power spectrum, and hence, 
for $\epsilon^2 \gtrsim 5 \times 10^{-9} (m_{e'}/100\,{\rm GeV})$, 
the abundance should satisfy $\Omega_{e'} h^2 \lesssim 10^{-3}$~\cite{Dolgov:2013una}. 
This can be another constraint on this scenario.

In this paper, in order to avoid the stable colored particles and the overproduced mirror electron,
let us consider another case where ${\cal Q}^q_{em}$ 
is given by
\begin{eqnarray}
\label{good scenario}
{\cal Q}^q_{em} &=& \tau^q_L+{\cal Q}^q_L +{\cal Q}^{R \, q}_{em},  \\
{\cal Q}^{R \, q}_{em}&=&\tau^q_R+ {\cal Q}^q_R, 
\end{eqnarray}
where ${\cal Q}^{R \, q}_{em}$ is the charge of $U(1)^R_{em}$, 
$\tau^q_R$ is the isospin given by the third component of $SU(2)_R$ 
and ${\cal Q}^q_R$ is the $U(1)_R$ charge. 
This breaking pattern is realized in a situation by introducing extra scalars charged 
under both $U(1)_{L}$ and $U(1)_{R}$.  
We introduce two such scalars denoted by $X_b$ and $X_l$ 
that are singlets under $SU(3)_c\times SU(2)_L \times SU(2)_R$, 
but have charges under both $U(1)_L$ and $U(1)_R$ as defined in Table \ref{table3}.

In our study, we consider two cases: 
\begin{enumerate}
\centering
\renewcommand{\labelenumi}{(\roman{enumi})}
\item[(I)] $\langle X_b \rangle =0$ and $\langle X_l \rangle  \neq 0$, 
\item[(II)] $\langle X_b \rangle \neq 0$ and $\langle X_l \rangle =0$.
\end{enumerate}
Note that $H_R$ also develops a nonzero VEV in both cases. 
The $U(1)^R_{em}$ gauge symmetry is broken by either $X_{b}$ or $X_{l}$, 
and a subgroup of $U(1)^R_{em}$ remains unbroken, 
similar to $Z^{em}_3$ discussed before.   
In the case (I), the unbroken symmetry is $Z^R_3$, 
while in the case (II) the unbroken symmetry is $Z^R_2$.
The scalar, $X_{b}$ ($X_{l}$), is charged under $Z^R_3$ ($Z^R_2$),
so that it is stable as far as its VEV is vanishing.
We note that the scalars are neutral under the EM symmetry
according to Eq.~(\ref{good scenario}) and 
the charge assignments in Table \ref{table3}. 
\footnote{If the ($U(1)_L, U(1)_R$) charge of $X_b$ is defined as ($-1/3, 1/3$), $X_b$
couples to down-type quarks at the renormalizable level and the phenomenology is
similar to the one discussed in Refs. \cite{Abe:2016wck,Okawa:2017bgh}. In this case, however,
we may suffer from the bound on the stable mirror up quark. }

\subsection{The interaction of the scalars  }

%%%%%%%%%%%%%%%%%%%%%%%%%%%%%%%%%%%%%%%%%%%%%%%%%%%%%%%%%%%
\begin{table}[t]
\begin{center}
\begin{tabular}{cccc}
\hline
\hline
Fields & spin    & ~~$U(1)_R$  & ~~$U(1)_L$   \\ \hline  
     $X_b$ & $0$       &            $-2/3$    &       $2/3$   \\
      $X_l$ & $0$      &            $1$    &       $-1$  \\ \hline \hline
\end{tabular}
\end{center}
\caption{The $U(1)_L \times U(1)_R$ charge assignment of the extra scalars. 
They are not charged under $SU(3)_c \times SU(2)_L \times SU(2)_R$.}
 \label{table3}
\end{table} 
%%%%%%%%%%%%%%%%%%%%%%%%%%%%%%%%%%%%%%%%%%%
We consider interactions between the additional scalars and the fermions.
The scalars, $X_b$ and $X_l$, 
are only charged under $U(1)_R \times U(1)_L$ 
as shown in Table \ref{table3}.
The charge assignments are defined to make the extra quarks and leptons unstable.
In this setup, the Yukawa couplings are written as: 
\begin{equation}
\label{Yukawa-X}
  {\cal L}_{Y}=- \lambda^{ij}_u \, X_b \, \overline{u^{ i}_R} \, u^{\prime \, j}_L -
   \lambda^{ij}_e \, X_l \, \overline{e^{ i}_R} \, e^{\prime \, j}_L  +h.c..
\end{equation}
The parity symmetry forces $\lambda^{ij}_u$ and $\lambda^{ij}_e$ to satisfy
\beq
\lambda^{ij}_u= \lambda^{ji \, *}_u,\quad \lambda^{ij}_e= \lambda^{ji \, *}_e.
\eeq
The extra fermions can decay to the SM fermions and $X_{b,l}$ 
through these Yukawa couplings.

Next, let us discuss the gauge interactions and the scalar potential. 
The Lagrangian involving $X_{b,l}$ is given by
\beq
{\cal L}_X = \sum_{\alpha=b,l} \left | \left (\pa_\mu - i g^\prime Q_\alpha  A^L_{\mu} + i g^\prime Q_\alpha A^R_{\mu} \right ) X_a \right |^2
-V_S,
\eeq
where $(Q_b,\,Q_l)=(2/3, \, -1)$ is defined.
$A^L_{\mu} $ and $A^R_{\mu}$ are the gauge fields of $U(1)_L$ and $U(1)_R$ symmetries, respectively. 
 $V_S$ is the scalar potential: 
\begin{eqnarray}
V_S&=&\Hat m^2_b|X_b|^2+\Hat m^2_l |X_l |^2 - m^2 \left (  |H_L|^2 + |H_R|^2 \right) \nonumber \\
&&+ \frac{\Hat \lambda_{b }}{2}|X_b|^4 + \frac{\Hat \lambda_{l }}{2} |X_l|^4+\Hat \lambda_{b l}|X_b|^2 |X_l|^2  \nonumber  \\
&&+ \left (\Hat \lambda^b_{H}|X_b|^2+\Hat \lambda^l_{H}|X_l |^2  \right ) \left (  |H_L|^2 + |H_R|^2 \right) \nonumber  \\
&&+ \frac{\Hat \lambda_+}{2} \left (  |H_L|^2 + |H_R|^2 \right)^2+ \frac{\Hat \lambda_-}{2} \left (  |H_L|^2 - |H_R|^2 \right)^2.
\end{eqnarray} 
The scalar potential $V_S$ induces the symmetry breaking, 
depending on the mass parameters in $V_S$.
We note that all the parameters in  $V_S$ can be defined as real valued,
so that there is no contribution to the $\theta$ term.
Conditions for a certain symmetry breaking is discussed in the next subsection.

Before discussing about the gauge symmetry breaking, 
we show the $U(1)_L \times U(1)_R$ gauge kinetic terms.
The kinetic terms of the gauge fields are
\beq
{\cal L}_{U(1)}=- \frac{1}{4} F^{\mu \nu}_L F_{L \, \mu \nu} - \frac{1}{4} F^{\mu \nu}_R F_{R \, \mu \nu}  - \frac{\epsilon}{2} F^{\mu \nu}_L F_{R\, \mu \nu },
\eeq
where $F^{\mu \nu}_L=\pa^\mu A_L^{\nu}-\pa^\nu A_L^{\mu}$ and $F^{\mu \nu}_R=\pa^\mu A_R^{\nu}-\pa^\nu A_R^{\mu}$ are defined.
$\epsilon$ is the kinetic mixing allowed by the parity symmetry.
To summarize, the parity transformation exchanges 
the gauge fields and the scalar fields as
\beq
A^L_{\mu} \leftrightarrow A^{R \, \mu}, \, H_L \leftrightarrow H_R, \, X_b \leftrightarrow X_b^\dagger,\, X_l \leftrightarrow X_l^\dagger.
\eeq
 
\subsection{The condition for the gauge symmetry breaking}

We study the vacuum structure given by $V_S$ and find out the condition for the parity and gauge symmetry breaking.
We expect that the VEVs of the scalars are not vanishing 
and each of them causes the corresponding symmetry breaking:
\begin{eqnarray}
\langle H_R \rangle \neq 0 &:& SU(2)_L \times SU(2)_R \times U(1)_L \times U(1)_R \to SU(2)_L  \times U(1)_L \times U(1)^R_{em},  \label{1}   \\
 \langle X_{b,l} \rangle \neq 0  &:&SU(2)_L  \times U(1)_L \times U(1)^R_{em} \to SU(2)_L  \times U(1)_Y , \label{2}  \\
\langle H_L \rangle \neq 0 &:&SU(2)_L  \times U(1)_Y \to U(1)_{em}. \label{3}
\end{eqnarray}
We discuss the symmetry breaking one by one. 
The scalar fields have the VEVs as 
\beq
\langle H_{R,L} \rangle =\frac{1}{\sqrt 2}  
\begin{pmatrix} 0 \\  v_{R,L} \end{pmatrix},\quad 
\langle X_{b,l} \rangle = \frac{1}{\sqrt{2}} \, v^{b,l}_X. 
\eeq
As mentioned above, the VEVs are real since all the parameters in $V_S$ are real.
The stationary condition for $H_R$ gives the equation for the VEVs: 
\beq
\label{stationary-H}
  \frac{\Hat \lambda_-}{2} (v^2_R-  v^2_L)+\frac{\Hat \lambda_+}{2} (v^2_L+  v^2_R) -m^2 + \frac{1}{2} \left (\Hat \lambda^b_{H} v^{b \,2}_X+\Hat \lambda^l_{H} v^{l \, 2}_X  \right )=0. 
\eeq
The VEV of $X_b$ ($X_l$) is vanishing, in the case (I) (case (II)). 

Let us focus on the case (I). 
Note that the results can be applied to the case (II) by replacing the index $b$ with $l$.
The stationary condition for $X_l$ is described as
\beq
\label{stationary-Xl}
 \frac{\Hat \lambda_l}{2} v^{l \, 2}_X +\Hat m^2_l + \frac{\Hat \lambda^l_{H}}{2} \,( v^2_R +v^2_L)=0. 
\eeq
In our setup, $v_L$ breaks the EW symmetry and is assumed to be tiny compared to the other VEVs. Then, assuming $v_L \ll v_R, \, v^{l }_X$, the stationary conditions, Eqs.~(\ref{stationary-H}) and (\ref{stationary-Xl}), lead an approximate condition for $v_R$ and $v^{l }_X$:
\beq
\begin{pmatrix} v^2_R  \\ v^{l \, 2}_{X} \end{pmatrix} \approx \begin{pmatrix}\Hat  \lambda  & \Hat \lambda^{l}_{H} \\
\Hat \lambda^{l}_{H}  &\Hat \lambda_l \end{pmatrix}^{-1} \begin{pmatrix} 2 m^2  \\ -2 \Hat m^{ 2}_{l} \end{pmatrix},
\eeq
where $\Hat \lambda =\Hat \lambda_+ +\Hat \lambda_-$ is defined.
When we choose the appropriate parameters in the right-hand side, we can realize
the symmetry breaking in Eqs.~(\ref{1}) and (\ref{2}).

Next, we discuss the EW symmetry breaking.
The Higgs boson $H_L$ should develop the non-vanishing VEV to 
cause the EW symmetry breaking.
After the parity symmetry is spontaneously broken down, 
the effective scalar potential is evaluated as
\begin{eqnarray}
\label{effectivepotential}
V^{eff}_S&=&m^2_b |X_b|^2+ \frac{\lambda_{b }}{2}|X_b|^4+( m^2_{eff} + \lambda^b_{H} |X_b|^2)   |H_L|^2 + \frac{\lambda}{2} \, |H_L|^4 .
\end{eqnarray} 
The parameters in the effective potential are renormalized, taking into account the corrections from $v^l_X$ and $v_R$. In particular, $m^2_{eff}$ is approximately evaluated as 
\begin{eqnarray}
m^2_{eff}&\approx&-\Hat \lambda_-  v_R^2,  
\end{eqnarray}
at the tree level. 
$m^2_{eff}$ is expected to be the source of the EW symmetry breaking, 
so that $\Hat \lambda_-$ should be tiny to obtain the EW symmetry breaking scale
that is much smaller than $v_R$. If $\Hat \lambda_-$ is vanishing, the global symmetry in $V_S$
is enhanced, so that the direction of $H_L$ could be interpreted as the pseudo-Goldstone boson.
In such a case, we may obtain the EW symmetry breaking radiatively \cite{Chacko:2005pe}.

In our work, we focus on phenomenology without specifying sources of the radiative correction to the scalar potential. 
We simply introduce a soft parity breaking term for $H_L$:
\beq
\Delta V= - \mu^2 |H_L|^2.
\eeq
Thus, the EW symmetry breaking scale is realized although we may have to allow fine-tuning for the Higgs mass term. This setup can evade the domain wall that would be generated by spontaneous parity breaking.

In the case (I), the VEV of $X_b$ is vanishing and the mass is given by $m^2_b$.
We note that the mass $m^2_b$ depends on $v_R$ and $v^l_X$ 
through the quartic couplings, namely $\Hat \lambda_{b l}$ and $\Hat \lambda^b_{H}$. 
Then, the mass of $X_b$ is expected to be around $v_R$ and/or $v^l_X$.
As will be shown in Sec. \ref{sec;phenomenology}, 
the VEV $v_R$ needs to be much higher than the EW scale 
to avoid experimental constraints. 
$X_b$ may reside at the very high-energy scale $\sim v_R$.  
The mass scale of $X_b$, however, also depends on other parameters in $V_S$. For simplicity, we study the phenomenology by treating $m^2_b$ as a free parameter.

\section{The solution to the strong CP problem}\label{sec;cp}
\label{CP problem}
We discuss the strong CP problem in this section. In general, the $\theta$ parameter is described effectively as
\beq
\overline \theta = \theta +\textrm{Arg}\left[\det(m_u m_d)\right],
\eeq
where $\theta$ is from the non-perturbative effect of the QCD vacuum and $m_{u,d}$ are the mass matrices for the SM quarks. 
The upper bound on $\overline \theta$ from the experiment is about $10^{-10}$~\cite{Baker:2006ts}. 
In our model, the parity symmetry is respected such that the $\theta$-term is forbidden. 
Note that the $\theta$-term explicitly breaks not only the CP symmetry 
but also the parity symmetry. This kind of scenario has been proposed motivated by the strong CP problem \cite{Babu:1989rb, Gu:2012in, Abbas:2017vle, Abbas:2017hzw, Gu:2017mkm}.

The parity symmetry is respected by introducing the mirror sector at some high scale 
until it is broken spontaneously.  
This means that the parity is broken to some extent at the low scale, e.g. the EW scale. 
An important fact of this model is that the Yukawa matrices for the mirror fermions 
are same as the SM ones at the high scale where the parity is conserved. 
The parity breaking scale, or equivalently the extra gauge symmetry breaking scale,  
should be higher than about $10^8$ GeV to make the first generation mirror fermions heavier than current experimental limits. 
Then, the RG correction to the $\theta$ term may be non-negligible  
as well as the corrections from the threshold and the higher-dimensional operators.

First, let us discuss how the $\theta$-term is vanishing near the parity breaking scale. 
In our model, the parity symmetry forbids the $\theta$-term at the tree level, namely $\theta=0$. 
Due to the existence of mirror particles, the quark matrices, $m_u$ and $m_d$, are replaced by
\begin{eqnarray}
m_u &\to & M_u= \begin{pmatrix} Y_u \dfrac{v_R}{\sqrt 2} & A_u \\  B_u   & Y^\dagger_u \dfrac{v_L}{\sqrt 2} \end{pmatrix}, \\
m_d &\to & M_d= \begin{pmatrix} Y_d \dfrac{v_R}{\sqrt 2} & A_d \\  B_d   & Y^\dagger_d \dfrac{v_L}{\sqrt 2} \end{pmatrix}, 
\end{eqnarray}
where $A_{u,d}$ and $B_{u,d}$ are $3 \times 3$ matrices. 
At the renormalizable level, $A_{u,d}$ and $B_{u,d}$ are vanishing in our model. 
Then, one can immediately realize
\beq
\label{Solution}
\det(M_{u})\propto \det(Y_{u}Y^\dagger_{u}),\quad 
\det(M_{d})\propto \det(Y_{d}Y^\dagger_{d}),
\eeq
which are both real numbers. Therefore, the $\overline \theta$ parameter, 
given by $\overline{\theta}= \theta + \textrm{Arg}\left[\det(M_u M_d)\right]$, is vanishing. 

The symmetry breaking may effectively generate nonzero $A_{u,d}$ and $B_{u,d}$.
In the case~(I), $X_b$ does not develop a VEV and the remnant symmetry is $Z^{R}_3$~\footnote{Actually, in this case it is an accidental global $U(1)$ symmetry which preserves mirror baryon number.}.  
Since the mirror quarks are charged under $Z^{R}_3$, 
the mass mixing terms, $A_{u,d}$ and $B_{u,d}$, are forbidden. 
Thus, the $\theta$-term is not generated even at low energy.
We give a discussion about loop corrections later.

In the case~(II), 
the VEV of $X_b$ is not vanishing, while that of $X_l$ is vanishing. 
The remnant symmetry is $Z^R_2$. 
The mirror leptons and the mirror down-type quarks are $Z^R_2$-odd 
while the mirror up-type quarks are $Z^R_2$-even. 
This implies that $A_{d}$ and $B_{d}$ are forbidden, but $A_{u}$ and $B_{u}$ are not.
In fact, the VEVs of the scalars generate  
\beq
B^{ij}_{u} = \lambda_u^{ij} \, \frac{v^b_X}{\sqrt{2}} ,\quad 
A^{ij}_{u} = \frac{a_u^{ij}}{\Lambda^2} \, v^b_X v_R v_L,
\eeq
where $\Lambda$ is a cut-off scale. If $\Lambda$ is very large,  $A_{u}$ is vanishing and 
$\overline{\theta}$ is also vanishing.
Otherwise, the Yukawa couplings, $ \lambda_u^{ij}$ and $a_u^{ij}$,
should be suppressed to evade the bound from the CP violation,
in the case (II) with $v^b_X \neq 0$.

Next, let us discuss the RG corrections.  
The relation in Eq.~(\ref{Solution}) is modified by the RG corrections that might revive the strong CP problem at low energy. The RG equations for the determinants of $Y_d$ and $Y_u$ are given by 
\begin{eqnarray}
\mu \frac{d}{d \mu} \det (Y_d)&=& \Tr (\beta_{Y_d} Y^{-1}_d) \det (Y_d), \label{RG1} \\
\mu \frac{d}{d \mu} \det (Y_u)&=& \Tr (\beta_{Y_u} Y^{-1}_u) \det (Y_u),\label{RG2}
\end{eqnarray}
where $\beta_{Y_d}$ and $\beta_{Y_u}$ are the $\beta$-functions defined as 
\begin{eqnarray}
\mu \frac{d}{d \mu} Y_d= \beta_{Y_d}, \quad 
\mu \frac{d}{d \mu} Y_u= \beta_{Y_u}.
\end{eqnarray}
If the right-hand sides of Eqs.~(\ref{RG1}) and (\ref{RG2}) are complex numbers 
and the imaginary parts of the determinants are amplified,  
a sizable $\theta$-term is predicted at the low scale. 
We note that the imaginary parts of $\det (Y_d)$ and $\det (Y_u)$ can be set to zero at the initial condition,
according to the phase rotation of quarks and the mirror quarks. 

Let us discuss the beta functions explicitly at the one-loop level.
After $H_R$ develops the non-vanishing VEV, we could integrate out the mirror fermions.
Then, the beta functions, $\beta_{Y_d}$ and $\beta_{Y_u}$, are evaluated as
\begin{eqnarray}
\beta_{Y_d}&=& \frac{1}{16 \pi^2} \left( -8 g^2_s- \frac{9}{4} g^2 - \frac{5}{12}g^{\prime 2} - \frac{3}{2} Y_u Y^\dagger_u + \frac{3}{2}  Y_d Y^\dagger_d +Y_2(H) \right) Y_d, \\
 \beta_{Y_u}&=& \frac{1}{16 \pi^2}  \left(  -8 g^2_s- \frac{9}{4} g^2 - \frac{17}{12}g^{\prime 2} + \frac{3}{2} Y_u Y^\dagger_u - \frac{3}{2} Y_d Y^\dagger_d +Y_2(H) \right) Y_u,
\end{eqnarray}
where $\gamma_H$ is given by
\beq
Y_2(H)=  3 \, \Tr (Y^\dagger_u Y_u +  Y^\dagger_d Y_d ) + \Tr (Y^\dagger_e Y_e).
\eeq
This leads the RG equations for $\det (Y_{u,d})$ as 
\begin{eqnarray}
\mu \frac{d}{d \mu} \ln( \det (Y_d) )&=& \frac{3}{16 \pi^2} \left( -8 g^2_s- \frac{9}{4} g^2 - \frac{5}{12}g^{\prime 2} - \frac{1}{2} \Tr(Y_u Y^\dagger_u) + \frac{1}{2} \Tr( Y_d Y^\dagger_d )+Y_2(H) \right), \nonumber \\
&&  \\
\mu \frac{d}{d \mu} \ln( \det (Y_u))&=& \frac{3}{16 \pi^2}  \left(  -8 g^2_s- \frac{9}{4} g^2 - \frac{17}{12}g^{\prime 2} + \frac{1}{2} \Tr( Y_u Y^\dagger_u )- \frac{1}{2} \Tr( Y_d Y^\dagger_d) +Y_2(H) \right). \nonumber \\
\end{eqnarray}
Thus, the imaginary parts of the left-handed sides of the RG equations are not evolved, 
since the right-handed sides are real at the one-loop level. In Appendix \ref{appendixA},
the relevant RG equations are summarized.

With the same spirit as in Ref.~\cite{Ellis:1978hq}, the contributions from renormalization of quark mass matrices alone are around $\mathcal{O}(10^{-16})$ even at the higher-loop level. 
The main differences are the existence of $X_b$, 
possible mixings in the Higgs sector and the kinetic terms of the $U(1)$ symmetry.  
These terms might lead non-vanishing corrections at loop levels.
In our model, the structure of chirality, the parity symmetry and the heavy masses of the mirror quarks however suppress the loop corrections.
Following Ref.~\cite{Ellis:1978hq}, the three-loop diagrams involving the $\lambda_u$ coupling
would lead a non-vanishing $\theta$ term in the case (I), while the one-loop diagram involving the CP-even scalars would contribute to $\overline \theta$ in the case (II) because of the non-vanishing VEV of $X_b$.
In both cases, their contributions are suppressed by $\lambda_u$ so that
the loop corrections would not spoil the tininess of $\bar{\theta}$ as far as the size of $\lambda_u$
is not too large. In fact, we assume that the alignment of $\lambda_u$ is unique to avoid the flavor constraints. Then, the loop corrections to $\overline \theta$ are much suppressed.

%%%%%%%%%%%%%%%%%%%%%%%%%%%%%%%%%%%%%%%%%%%%%%%%%%%%%%%%%%%%%%%%%%%%%%%%%%%%%%%%%%%%%%%%%%%%%%%%%%%%%%%%%%%%%%%%%%%%%%%%%%

\section{Phenomenology}
\label{sec;phenomenology}

We study the phenomenology in this section.
In our model, there are Yukawa couplings involving the scalars, 
the mirror fermions and the SM fermions:
\beq \label{eq;lambda}
\lambda^{ij}_\psi  \overline{\psi^{ i}_R}X \psi^{\prime \, j}_L  +h.c. ~(\psi=u, \,e),
\eeq
as shown in Eq.~(\ref{Yukawa-X}). 
Because of the parity symmetry, 
$\lambda^{ij}_\psi$ is a hermitian matrix and 
the mirror fermion mass ratios are the same as the SM predictions 
above the parity breaking scale.
The mirror up quark and electron are the lightest mirror quark and lepton 
that are expected to dominantly contribute to the low-energy physics. 
Note that $\lambda^{ij}_\psi $ is defined in the mass base. 
Our main motivation of this paper is to study physics involving our DM candidates.
In particular, we will numerically analyze the parameter region allowed by the DM physics
and discuss the flavor and LHC physics relevant to the result.
We study the phenomenology of the case (I) in Subsection~\ref{sec;XbDM} 
and of the case (II) in Subsection~\ref{sec;XlDM}.

\subsection{Case (I) : baryonic DM ($X_b$) scenario}\label{sec;XbDM}
We first consider the case that $X_b$ does not develop the VEV.
In this case, the $Z^R_3$ symmetry remains unbroken 
as the remnant of the subgroup of $U(1)^R_{em}$,  
and it makes $X_b$ stable. 
The scalar $X_b$ couples to the SM up-type quarks and the mirror quarks 
via the $\lambda_u$ coupling. 
Since $X_b$ is stable and couples with the SM particles 
involving the mirror up quark, the coupling leads a suitable co-annihilation cross section and
$X_b$ can be a good DM candidate. 
We focus on the Yukawa interactions involving the mirror up quark in phenomenology. 
Furthermore, we consider the following three cases that $X_b$ dominantly couples to 
\begin{enumerate}
\renewcommand{\labelenumi}{(\Alph{enumi})}
\item up quark ( $|\lambda^{uu'}_u| \gg |\lambda^{cu'}_u| ,\, |\lambda^{tu'}_u| $), 
\item charm quark ( $|\lambda^{cu'}_u| \gg |\lambda^{uu'}_u| ,\, |\lambda^{tu'}_u| $), 
\item top quark ( $|\lambda^{tu'}_u| \gg |\lambda^{uu'}_u| ,\, |\lambda^{cu'}_u| $). 
\end{enumerate}
We discuss the predictions and constraints of each case below. 
Note that contours for these parameters in each case should be interpreted as upper bounds since in principle these couplings could be present simultaneously.  
In the Higgs-portal DM scenario, there are many discussions in the literature~\cite{Kanemura:2010sh,Djouadi:2011aa,Djouadi:2012zc,Escudero:2016gzx,Balazs:2017ple,Athron:2018ipf}, so we shall not repeat here.
In Appendix \ref{appendixB}, the contribution of the Higgs portal coupling between DM and
the SM Higgs is summarized and the impact on our result is shortly discussed.

\subsubsection{Dark Matter Physics}
To begin with, we discuss DM relic abundance, direct detection and indirect detection in the $X_b$ DM scenario, 
assuming that $X_b$ was thermally produced in our universe. 
These observations put constraints on the Yukawa couplings, $\lambda^{ij}_u$ 
and masses of the DM and the mirror fermion.  
We first study the general features in this kind of model, 
and then elaborate on each case listed above.

This DM candidate mainly annihilates into a pair of the up-type quarks
by exchanging the mirror quarks in the $t$-channel. 
We assume that flavor violating annihilations, 
$X_b X_b^\dagger \to u_i \bar{u}_j$ ($i \neq j$), are negligibly small,   
and the dominant processes are flavor conserving annihilations. 
In the non-relativistic region where the thermal freeze-out occurs, 
the cross section can be expanded in terms of the relative velocity, $v$, of incoming DM particles, 
\begin{equation}
(\sigma v)_{X_b X_b^\dagger \to u_i \bar{u}_i} = a_i + b_i v^2 ,
\end{equation}
where 
\begin{align}\label{eq:anncs}
a_i & = \frac{N_c m_i^2 }{16\pi} 
\left[ \sum_{j} \frac{ |\lambda_u^{ij} |^2 }{ {m^\prime_j}^2 + m_X^2 } \right]^2 ,\\
b_i & = \frac{N_c m_X^2 }{48 \pi} 
\left[ \sum_{j} \frac{ |\lambda_u^{ij}|^2 }{ {m^\prime_j}^2 + m_X^2 } \right]^2 ,
\end{align}
in the massless quark limit ($m_i \ll m^\prime_j, m_X$). 
$m_X$ denotes the DM mass. 
$m_i$ and $m^\prime_j$ are masses of the SM and mirror up-type quarks: 
$(m_1, \, m_2, \, m_3)=(m_u, \, m_c, \, m_t)$ and 
$(m^\prime_1, \, m^\prime_2, \, m^\prime_3)=(m_{u^\prime}, \, m_{c^\prime}, \, m_{t^\prime})$, respectively. 
The partial $s$-wave of this process is suppressed by a quark mass in the final state. 
Thus, the pair annihilation will be $p$-wave dominant except for the top-philic case, (C). 
In the other cases, (A) and (B), the large Yukawa couplings are required to achieve the observed relic abundance. 

The Yukawa couplings, $\lambda_u^{ij}$, also give rise to elastic DM-nuclei scattering. 
The DM-quarks effective interaction relevant for the spin-independent (SI) scattering is given by 
\begin{equation}
\begin{split}
{\cal L}_{\rm eff} = & \sum_q \left[ C_{S,q} m_q X_b^\dagger X_b \overline{q} q 
+ C_{T,q} (\partial_\mu X_b^\dagger \partial_\nu X_b) {\cal O}_{T,q}^{\mu\nu} \right] \\
& + C_{S,g} X_b^\dagger X_b \frac{\alpha_s}{\pi} G_{\mu\nu} G^{\mu\nu} 
+ \sum_q C_{V,q} (i X_b^\dagger \overleftrightarrow{\partial_\mu} X_b) \overline{q} \gamma^\mu q ,
\end{split}
\label{eq;LeffDD}
\end{equation}
where we define $\phi_2 \overleftrightarrow{\partial_\mu} \phi_1 
\equiv \phi_2 \partial_\mu \phi_1 - \partial_\mu \phi_2 \cdot \phi_1$ 
and the twist-2 operator, 
\begin{equation}
{\cal O}_{T,q}^{\mu\nu} \equiv \frac{i}{2} \bar{q} \left( \gamma^\mu \partial^\nu + \gamma^\nu \partial^\mu - \frac{1}{2} g^{\mu\nu} \Slash{\partial} \right) q .
\end{equation}

\begin{figure}[!t]
	\begin{center}
		{\epsfig{figure=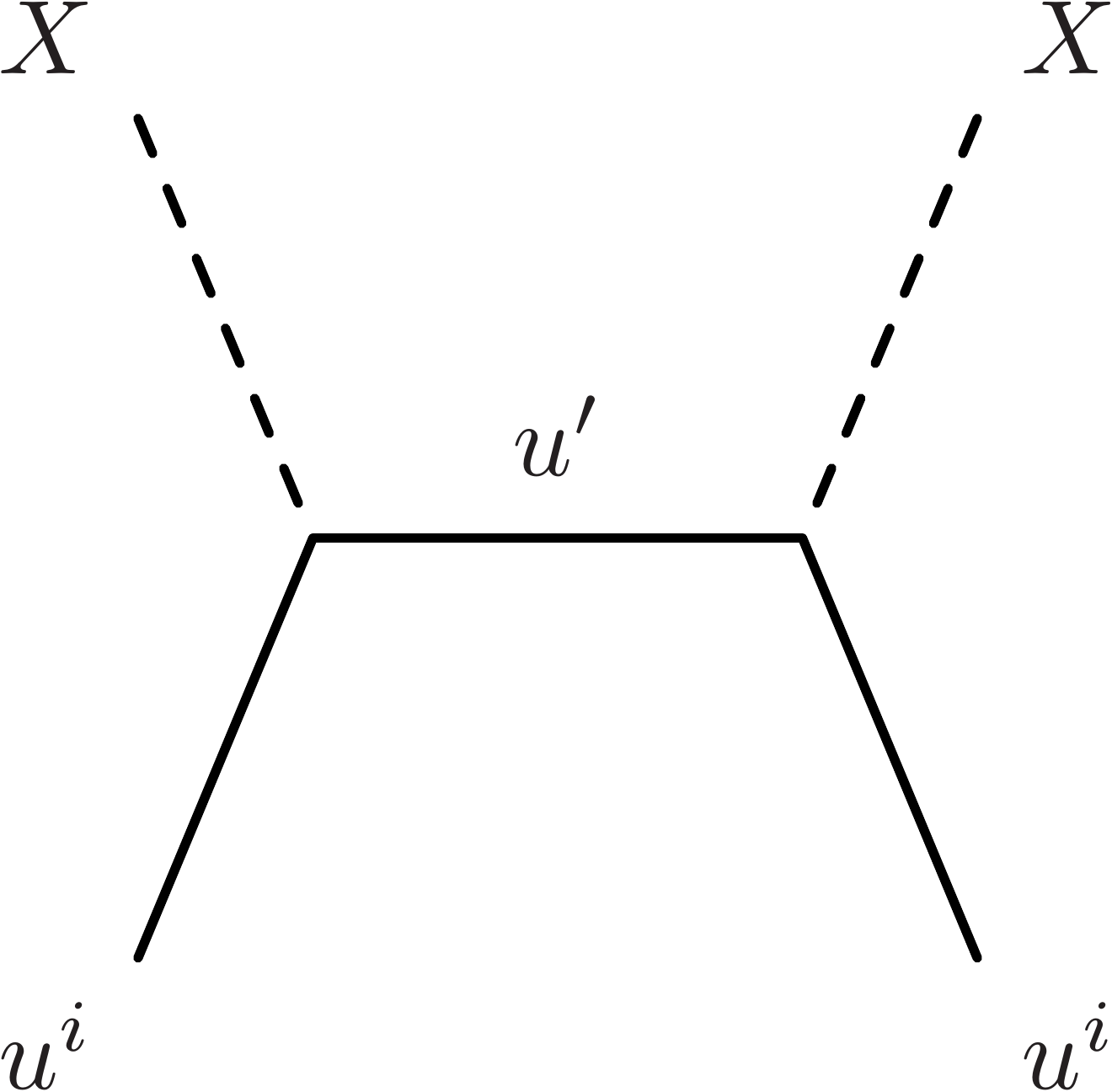,width=0.25\textwidth}}\hspace{1cm}{\epsfig{figure=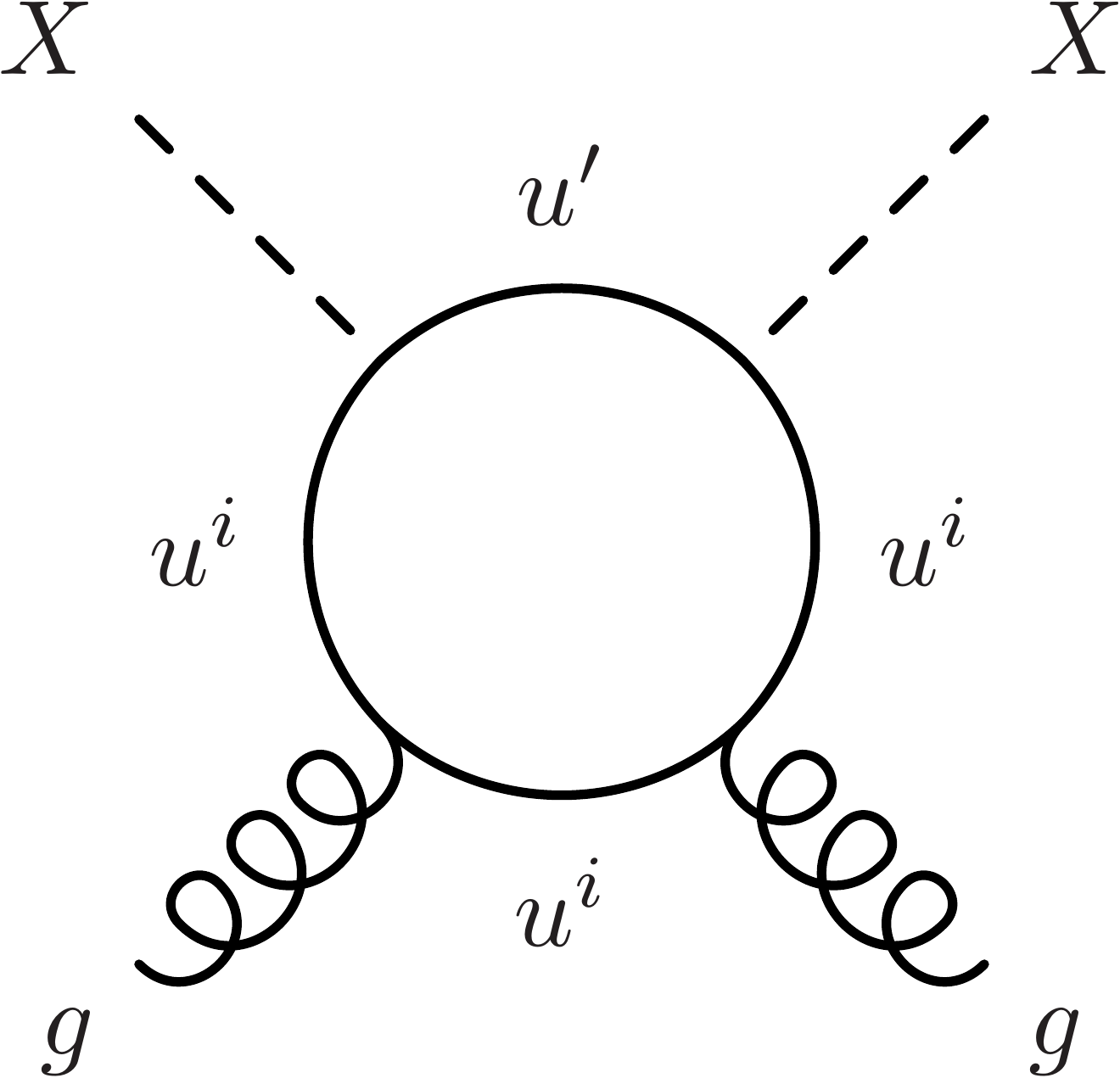,width=0.25\textwidth}}\hspace{1cm}{\epsfig{figure=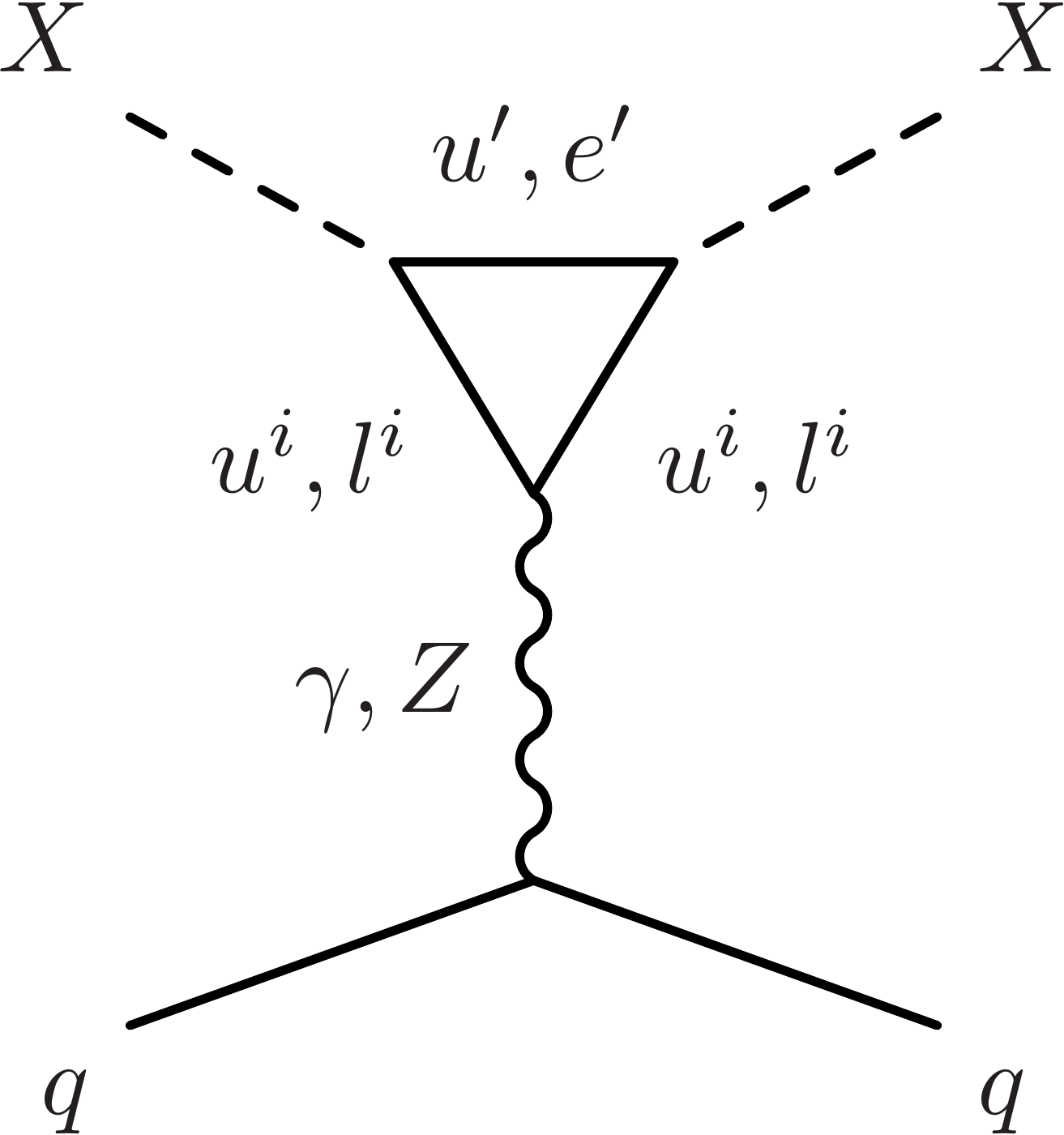,width=0.25\textwidth}}
		\caption{Example diagrams relevant for DM-nucleus elastic scattering. \label{fig;direct} } 
	\end{center}
\end{figure}

At the tree level, the mirror fermion exchanging, as shown in the left panel of Fig.~\ref{fig;direct}, generates $C_{S,q}$, $C_{T,q}$ and $C_{V,q}$: 
\begin{align}
C_{S, u_i} &= \sum_j \frac{|\lambda_u^{ij}|^2}{4} 
\frac{ 2 m^{\prime 2}_j - m_X^2 }{ ( m^{\prime 2}_j - m_X^2 )^2 } , \\
C_{T, u_i} &= \sum_j \frac{ |\lambda_u^{ij}|^2 }{ ( m^{\prime 2}_j - m_X^2 )^2 } ,\\
C_{V, u_i} &= - \sum_j \frac{ |\lambda_u^{ij}|^2 }{ 4 (m^{\prime 2}_j - m_X^2) } ,
\end{align}
where we neglect the SM quark masses. 
The DM particles can also scatter off the gluon in the nucleon 
via box diagrams shown in the central panel of Fig.~\ref{fig;direct}. 
Integrating out the short-distance contribution, we find the coefficient to be 
\begin{equation}
C_{S, g} = \sum_{i,j} \frac{ |\lambda_u^{ij}|^2 }{ 24 (m^{\prime 2}_j - m_X^2) } .
\end{equation}
Note that this equation is valid only when the mirror quarks and the DM are 
sufficiently heavier than the SM quarks. 
For the result that keeps the quark masses finite, see e.g. Ref.~\cite{Hisano:2015bma}. 

There are important loop processes as well. 
At the one-loop level, photon and $Z$ boson can mediate the DM-nuclei scattering via penguin diagrams as shown in Fig.~\ref{fig;direct}. 
The photon exchanging induces the DM coupling to the quark vector current with the coefficient, 
\begin{equation}
C_{V, q}^\gamma = \frac{\alpha Q_u Q_q N_c}{4\pi} 
\sum_{i,j} \frac{|\lambda_u^{ij}|^2}{m_X^2} 
I_1 (m^{\prime 2}_j/m_X^2, m_i^2/m_X^2) ,
\label{eq:photon_exchange}
\end{equation}
where 
\begin{equation}
\begin{split}
I_1(x,y) = \frac{1}{3} \int^1_0 dt &\left[ t^3 (2t-3) \left( \frac{1}{D(x,y)} - \frac{1}{D(y,x)} \right) \right. \\
&\left. -\ t^4 \left( \frac{x + (1-t)^2}{2D(x,y)^2} - \frac{y + (1-t)^2}{2{D(y,x)}^2} \right) \right] ,
\end{split}
\end{equation}
with 
\begin{equation}
D(x,y) = t (1-t) - x t - y (1-t) .
\end{equation}
Similarly, the contribution from the $Z$ boson exchanging is evaluated as 
\begin{equation}
C_{V, q}^Z = \sqrt{2} G_F N_c \frac{g_{V,q}}{16\pi^2} \sum_{i,j} |\lambda_u^{ij}|^2
\frac{m_i^2}{m_X^2} I_2 (m^{\prime 2}_j/m_X^2, m_i^2/m_X^2) ,
\label{eq:Z_exchange}
\end{equation}
where $g_{V,q} = (T_3)_q - 2 Q_q \sin^2\theta_W$ and 
\begin{equation}
I_2 (x,y) = \int^1_0 dt\, \frac{ (1-t)^2}{D(x,y)}.
\end{equation}
In the limit that $m^\prime_j \gg m_i, m_X$, Eqs.(\ref{eq:photon_exchange}) and (\ref{eq:Z_exchange}) reduce to simpler forms, 
\begin{align}
C_{V,q}^\gamma &= \frac{\alpha Q_u Q_q}{4\pi} \frac{N_c}{3} 
\sum_{i,j} \frac{|\lambda_u^{ij}|^2}{m^{\prime 2}_j} 
\ln\left( \frac{m_i^2}{m^{\prime 2}_j} \right) ,\\
C_{V,q}^Z &= \sqrt{2} G_F N_c \frac{g_{V,q}}{16\pi^2} 
\sum_{i,j} |\lambda_u^{ij}|^2 \frac{m_i^2}{m^{\prime 2}_j} 
\ln \left( \frac{m_i^2}{m^{\prime 2}_j} \right) .
\end{align}
We find that the $Z$-exchanging contribution is proportional to the quark mass squared, 
and then it will be significant only in the top-philic case. 

In the following discussion, we only keep contributions from the mirror up quark, 
and discuss the cases (A), (B) and (C) 
where the mirror up quark dominantly couples to the up, charm and top quark, 
respectively. 
The DM candidate, $X_b$ in this subsection and $X_l$ in the next subsection, 
is simply denoted as $X$.

%-----------------
%------------------------------

\subsubsection*{(A) Up-philic case}

In this case, the DM particle $X$ 
scatters off the valence up quark in nucleons at the tree level. 
The up-philic coupling is strongly constrained from direct detection experiments. 

Let us estimate the elastic DM-nucleon scattering cross section, 
assuming the DM abundance was produced via this coupling. 
From Eq.~(\ref{eq:anncs}), the DM pair annihilation cross section is given by 
\begin{equation}
\sigma v \simeq \frac{|\lambda_u^{uu'}|^4}{16 \pi} \frac{m_X^2}{m_{u'}^4} v^2 ,
\end{equation}
with good approximation. 
This formula shows how $\lambda_u^{uu'}$, $m_X$ and $m_{u'}$ are related 
to each other via the observed abundance. 
Using an approximate solution for the thermal relic abundance, 
$\Omega h^2 \simeq 0.12 \times 10^{-36}\mbox{cm}^2 / \langle \sigma v \rangle$, 
the SI scattering cross section is approximately given by 
\begin{equation}
\sigma_{SI} \simeq \left(\frac{1\,\mbox{TeV}}{m_X}\right)^2 \times 10^{-41} \, [\mbox{cm}^2] .
\end{equation}
This is several orders of magnitude larger than the current XENON1T bound~\cite{Aprile:2018dbl}, 
and then we conclude that the up-philic case has already been excluded.

%-----------------
\begin{figure}[!t]
	\begin{center}
		{\epsfig{figure=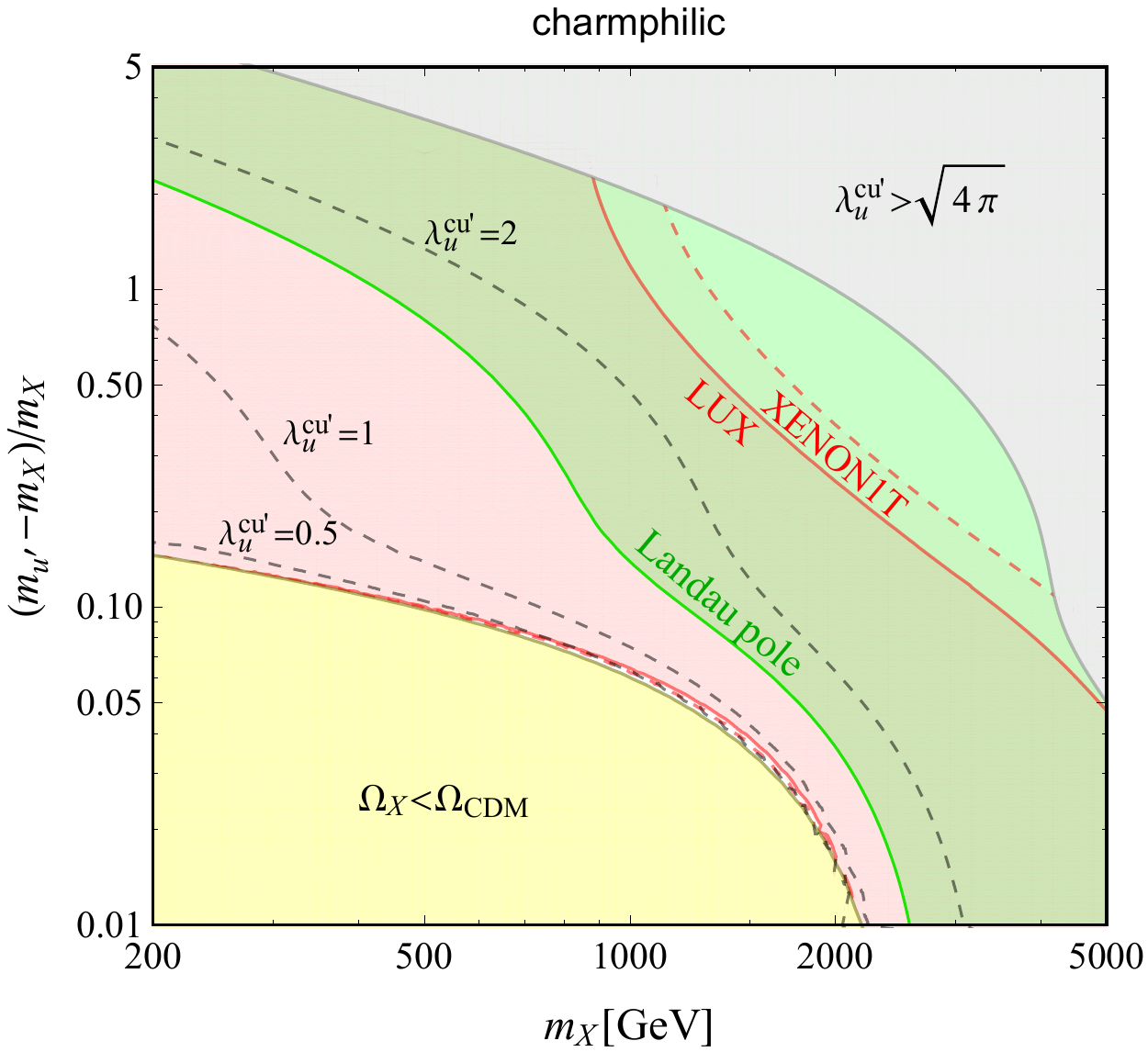,width=0.45\textwidth,height=0.42\textwidth}}{\epsfig{figure=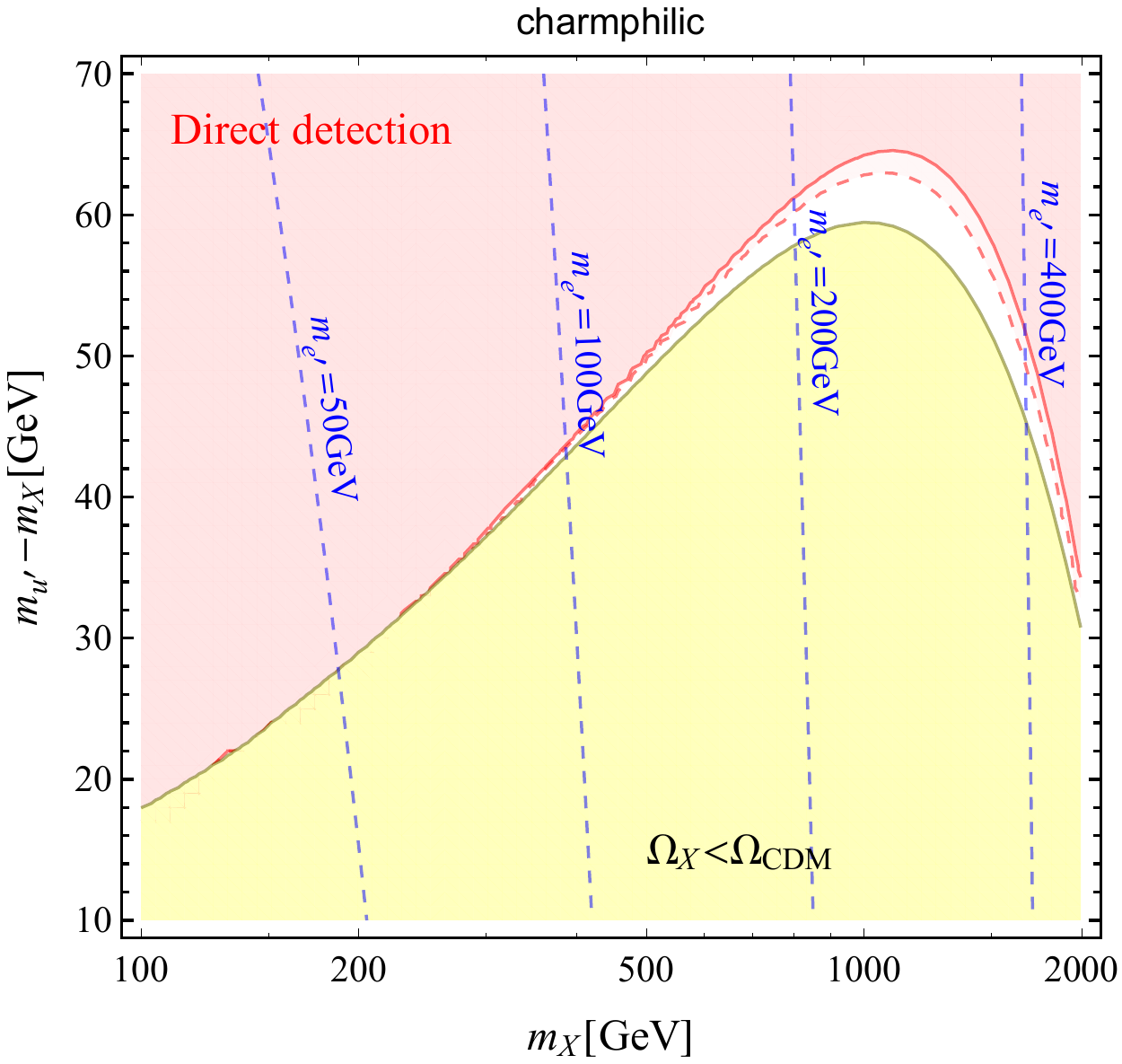,width=0.45\textwidth,height=0.41\textwidth}}
		\caption{ $m_X$ vs. the mass difference between $u^\prime$ and $X$ in the charm-philic case. Black dashed lines in the left panel show the contours with $\lambda^{cu^\prime}_{u}=0.5,1,2$, while blue dashed lines show the corresponding mirror electron's mass. White region in the right panel can satisfy relic abundance and evade various constraints. \label{fig;charm} } 
		\end{center}
\end{figure}
%---------------
\subsubsection*{(B) Charm-philic case}

The charm-philic case is similar to the up-philic one in the DM annihilation, 
so a large Yukawa coupling is required for the relic abundance. 
However, since the charm quark is a sea quark in nucleons, 
the tree-level process does not generate the couplings of the DM to nucleon vector current and 
thus the  SI cross section is much smaller than the up-philic case. 
In this case, the dominant contribution comes from the one-loop photon exchanging 
in most parameter space. The exception is a compressed region, $m_X \simeq m_{u'}$, 
where the DM-gluon scattering via the box diagrams can dominate over the former contribution. 

In Fig.~\ref{fig;charm}, we show how various contours in the parameter space are confronted with relic abundance, direct detection and perturbativity. The yellow area gives $\Omega_X< \Omega_{\textrm{CDM}}$: the coannihilation processes are too efficient and the produced DM abundance is below the observed one even for $\lambda_u^{cu'} = 0$\footnote{In fact, $\lambda_u^{cu'}$ cannot be vanishing and has to be large enough for $Xg \leftrightarrow u' \bar{u}^i$ process to frequently occur at freeze-out. In our case, the condition is fulfilled for $\lambda_u^{cu'} \gtrsim 10^{-4}$.}. The pink regions are excluded by LUX (solid)~\cite{Akerib:2016vxi} and XENON1T (dashed)~\cite{Aprile:2018dbl} experiments. On the left panel, regions above the green contour labeled with Landau pole would give too big couplings below the $W_R$ boson mass scale, part of which has been already constrained by the LUX and XENON1T experiments. 
The detail of our analysis on this bound is shown in Appendix \ref{appendixA}. The gray region satisfies $\lambda_{u}^{cu'}>\sqrt{4\pi}$. 
As a result, only the white region in the right panel can satisfy the observed DM abundance and can escape the various constraints. 
The blue dashed lines on the right panel show the values of the mirror electron mass translated from the mirror up quark mass using $m_{e'} \simeq (m_e/m_u) \, m_{u'}$.

%-------------
\begin{figure}[!t]
	\begin{center}
		{\epsfig{figure=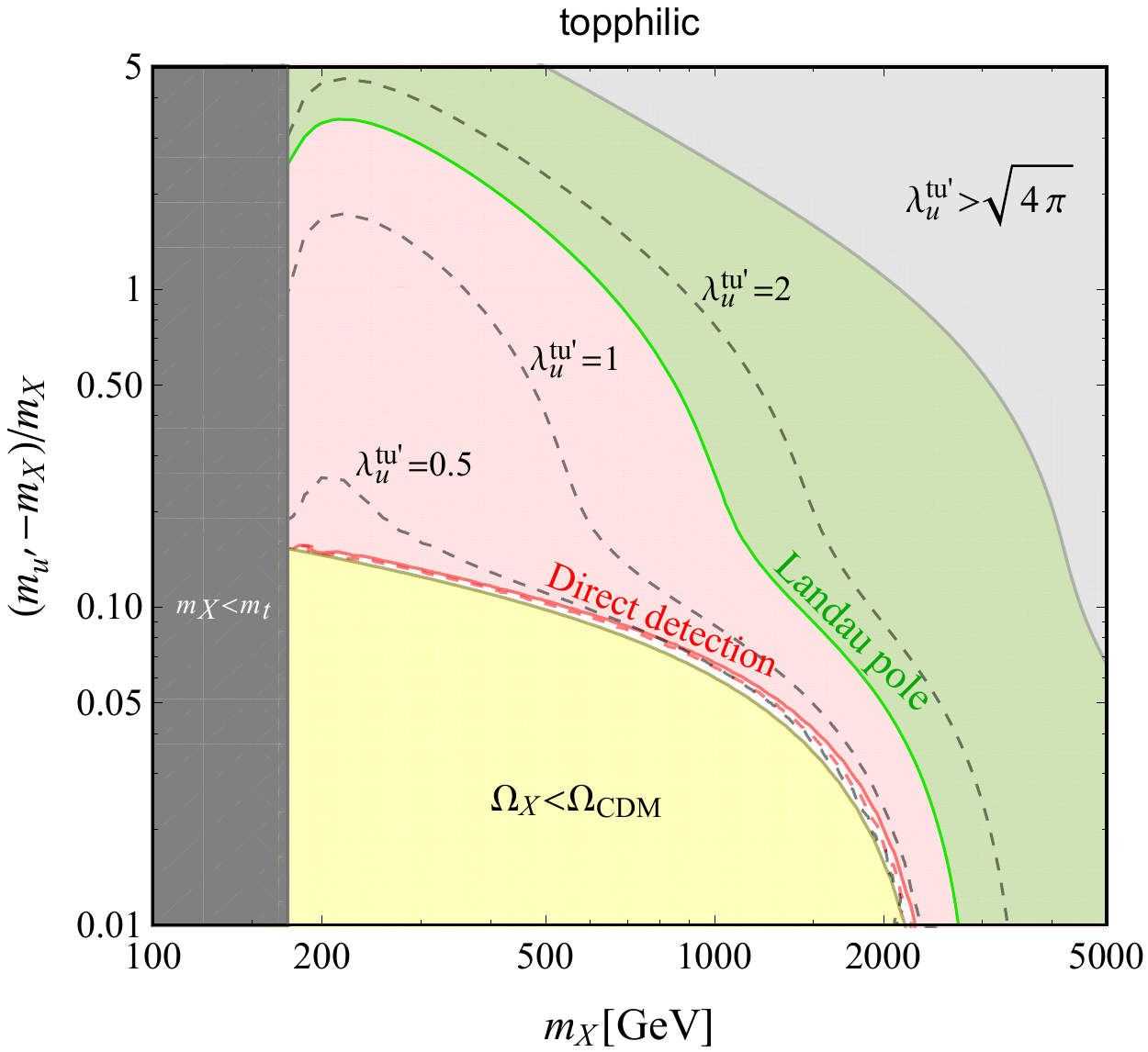,width=0.45\textwidth,height=0.42\textwidth}}{\epsfig{figure=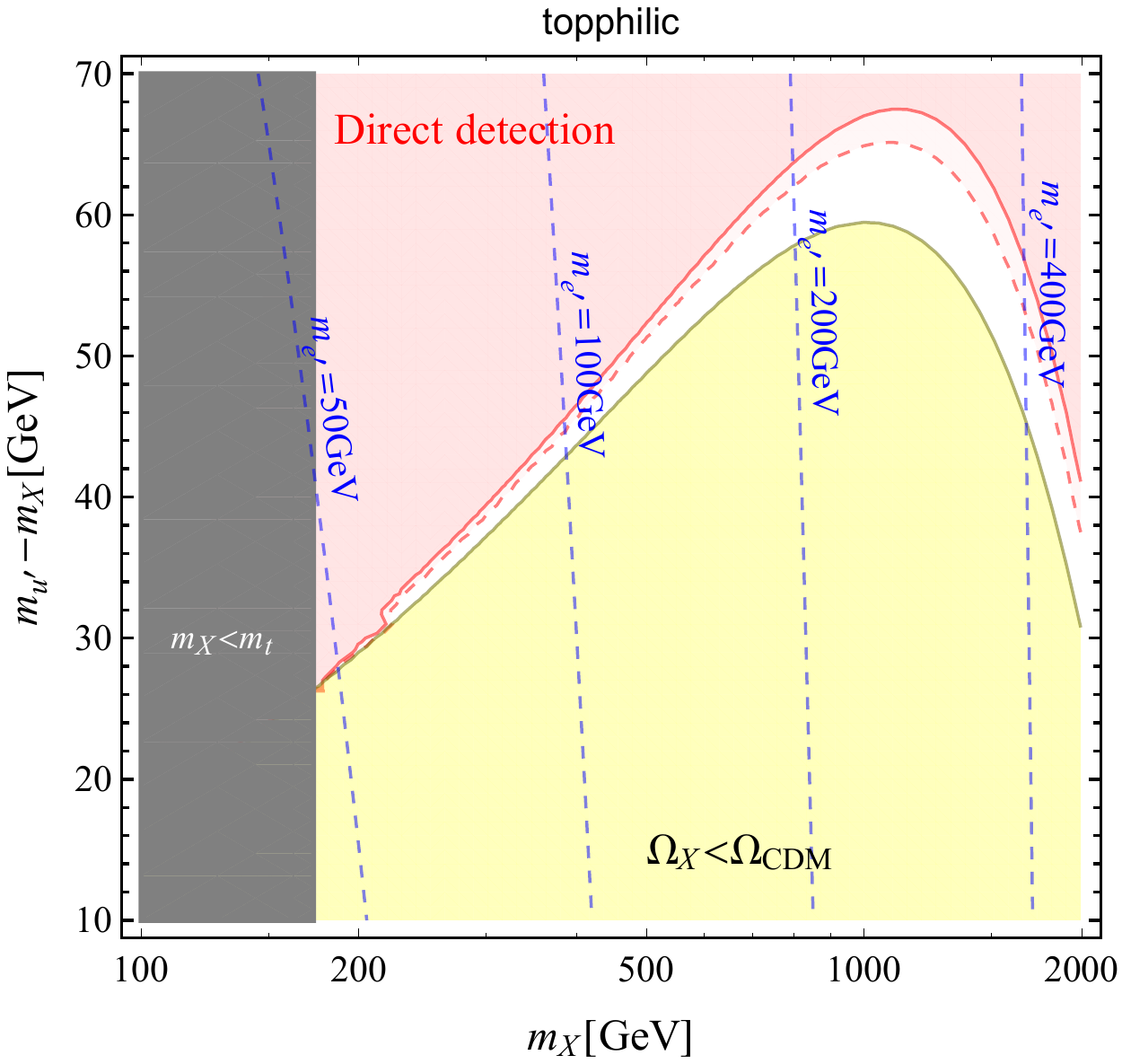,width=0.45\textwidth,height=0.41\textwidth}}
		\caption{Similar to the charm-philic case, $m_X$ vs. the mass difference between $u^\prime$ and $X$ in the top-philic case. White region is relatively larger than that in charm-philic case. \label{fig;top}}
	\end{center}
\end{figure}
%------------------------------

\subsubsection*{(C) Top-philic case}

The top-philic case is more complicated than the other two cases. 
Since the top quark is much heavier than other quarks, 
the $s$-wave contribution is not so suppressed in $X X^\dagger \to t \bar{t}$. 
Then, a smaller Yukawa coupling, $\lambda_u^{tu'}$, is predicted in this case. 
In direct detections, the $Z$ exchanging is the dominant contribution, instead of the photon exchange process, because of the large top mass. 
Figure~\ref{fig;top} shows various contours and constrains, similar to the charm-philic case. 
We can see that the white region on the right panel is a bit larger than that in the charm-philic case. 
Note that models of the DM with a top partner have also been studied in other literatures, e.g, in Refs.~\cite{Baek:2016lnv,Blanke:2017tnb}.

We point out that in all the cases the indirect searches from cosmic rays, 
gamma-ray and neutrinos do not pose any pressing limit, due to some suppressions. 
The tree-level process, $XX^\dagger \to q \overline{q}$, is $p$-wave suppressed. 
In the early universe when DM was freezing out, the velocity was about $1/3$, 
while at the present time $v\sim 10^{-3}$. Therefore the annihilation cross section at present is 
$10^{-6}$ times smaller than the canonical value for the thermal relic. 
Besides, by closing the external quark lines, $X X^\dagger \to gg$ is obtained at the one-loop level. 
This is $s$-wave dominant and not suppressed by small $v$, and then it may dominate the DM annihilation at the present. It is, however, loop-suppressed  by a factor, $\alpha_s^2/(4\pi)^2$. In the parameter space considered above, the annihilation cross section is at most 
$4\times10^{-28}\, {\rm cm^3/s}$ that is $10^{-2}$ times smaller than the canonical value for thermal relic. 
Thus, we conclude that the indirect searches have little impact on our model.

\subsubsection{Flavor Physics}
\label{sec;flavor1}

In our study of flavor physics, we assume that only one element of $\lambda^{ij}_u$ is sizable
and the others are negligibly small. 
The processes involving the lightest mirror quark, $u^\prime$,
are the most sensitive ones to the physical observables at low energy.
Then, the relevant Yukawa couplings between the mass eigenstates of the fermions 
and $X_b$ are 
\beq
\label{eq;relevantYukawa}
\lambda^{uu'}_u \, \overline{u_R} \, X \, u^\prime_L + \lambda^{cu'}_u \, \overline{c_R} \, X \, u^\prime_L+ \lambda^{tu'}_u \, \overline{t_R} \, X \, u^\prime_L+h.c..
\eeq
In the DM physics, we investigated the three cases:
(A) up-philic case, (B) charm-philic case, and (C) top-philic case.
In general, constraints from the flavor physics are very tight, 
even though the new physics scale is much higher. 
In this subsection, we discuss the constraint from flavor physics relevant to the DM physics
in each case, taking into account the small Yukawa couplings
irrelevant to the DM physics as well.

In the baryonic DM scenario, namely the case (I), the DM candidate $X_b$ interacts with the up-type quarks via the Yukawa couplings.
There are also gauge interactions induced by the $Z^\prime$ and the $Z$-$Z^\prime$ mixing, but
they are suppressed by the large $Z^\prime$ mass.
The dominant contribution to flavor observables is effectively induced 
by the Yukawa interactions in Eq.~(\ref{eq;relevantYukawa}).

First, let us discuss the cases (A) and (B).
In those setups, either $|\lambda^{uu'}_u|$ or $|\lambda^{cu'}_u|$
is large. If the other element is also sizable, flavor violating couplings would
be effectively generated by integrating out the mirror quark and $X_b$. For instance,
if $|\lambda^{cu'}_u|$ ($|\lambda^{uu'}_u|$) is not vanishing in the case (A)
(in the case (B)), the four-fermion coupling that contributes to the $D$-$\overline{D}$ mixing
is generated at the one-loop level. Such a $\Delta F=2$ process is generally most sensitive to new physics, 
so that we numerically estimate the bound on $\lambda^{ij}_u$ below.

The effective operator that contributes to the $D$-$\overline{D}$ mixing
is generated by the box diagram involving the mirror quarks and $X_b$:
\beq
{\cal H}_{eff}= C_D \, \left ( \overline{u_{R}} \gamma_\mu c_R \right ) \left ( \overline{u_{R}} \gamma^\mu c_R \right ),
\eeq
where $C_D$ is evaluated at one-loop level as
\beq
C_D = \lambda^{ui}_u \lambda^{ci \,*}_u \lambda^{uj}_u \lambda^{cj \,*}_u \, \frac{1 }{64 \pi^2}   \frac{1}{m^{ \prime 2}_{i}-m^{ \prime 2}_{j}}
\left \{ x_i f(x_i) -x_j f(x_j) \right \}.
\eeq
$x_i$ is defined as $x_i\equiv m^{\prime 2}_i/m^2_X$ and
$f(x)$ is the function satisfying 
\beq
f(x)= \frac{x}{(x-1)^2} \ln x -\frac{1}{x-1}.
\eeq
In the cases (A) and (B), $C_D$ is approximately estimated as
\beq
C_D \approx \frac{ \left ( \lambda^{uu'}_u \lambda^{cu' \,*}_u \right )^2 }{192 \pi^2}   \frac{1}{m^2_X},
\eeq
assuming $m_{u^\prime} \approx m_X$. 

One relevant observable concerned with the $D$-$\overline{D}$ mixing is 
the mass difference described as
\beq
\Delta M = \frac{2}{3} \, |C_D| \, \eta_D m_D f^2_D \Hat B_D.   
\eeq
The parameters on the right-hand side are numerically known as
 $m_D=1864.83 \pm 0.05$ MeV,
$f_D=212.15 \pm 1.45$ MeV,
$\eta_D=0.772$, and $\Hat B_D=0.75 \pm 0.02$ \cite{PDG,HFLAV}.

The value of $\Delta M $ is measured by experiments, and should be small enough
to evade the experimental bounds. For instance,
in Ref. \cite{HFLAV}, the measured value of $\Delta M$ 
is $0.04 \, \% \leq \Delta M \, \tau_D  \leq 0.62 \, \%$ at 95 \% CL. 
If we require the new physics contribution to be less than $0.1 \, \%$, we
can obtain the bound on $\lambda^{ij}_u$ in the cases (A) and (B) as
\beq
 \left |\lambda^{uu'}_u \lambda^{cu'}_u \right | \lesssim 0.005~(0.01),
\eeq
when $m_{u^\prime} \approx m_X$ is imposed and $m_X$ is fixed at 500 GeV (1 TeV).

In the case (C), on the other hand, the strong bound from the meson mixing becomes much milder, since $u^\prime$ dominantly couples to the top quark. Instead, 
the exotic top decay in association with a gauge boson in the final state might severely constrain our model.
The current experimental upper bound on such a process is given 
in Ref. \cite{PDG}, roughly ${\cal O}(10^{-4})$.
For instance, 
the flavor-violating top decay to a photon or gluon and one light quark is induced by the following operator generated at the one-loop level:
\begin{eqnarray}
{\cal L}_t &=& C^{ it }_{u} \left [ \frac{2}{3} \,  \frac{e}{16 \pi^2}\,m_{t} \, \overline{u^i_R} \, \sigma_{\mu \nu} t_L  \, F^{\mu \nu}\right ] + C^{ it }_{u} \left [ \frac{g_s}{16 \pi^2} \, m_{t} \, \overline{u^i_R} \, \sigma_{\mu \nu} t^a t_L  \, G^{a \, \mu \nu}\right ], 
\end{eqnarray}
where $\sigma_{ \mu \nu}= \frac{i}{2} [ \gamma_\mu, \gamma_\nu ]$ is defined. 
$F^{\mu \nu}$ and $G^{a \, \mu \nu}$ are the gauge field strengths that consist of photon and gluon, respectively.
$ C^{ ij }_{u }$ is given by 
\beq
 C^{ ij }_{u}=  - \frac{\lambda^{ik}_u \lambda^{jk \, *}_u }{24  \, m_X^2}\, f_7 (x_k),
\eeq
where $f_7(x)$ is defined as
\beq
f_7(x) = \frac{1}{ (x-1)^4} \left \{ 2+ 3 x -6x^2 +x^3 + 6 x \ln x  \right \}.
\eeq
Using the coefficients, each partial decay width of the flavor-violating top decays can be estimated. In particular, the decay to a light quark and gluon is larger than the others, because of the relatively large gauge coupling. 
We conclude that our prediction of the branching ratio is
less than $10^{-6}$ and negligible for the current experimental bound, even if the Yukawa couplings are ${\cal O}(1)$ and the DM mass is ${\cal O}(100)$ GeV.
In our model, the flavor-violating top decay associated with a $Z$ boson is also possible,
but the prediction is also much below the current experimental bound.
Eventually, the strongest bound on our model in the case (C) comes from the direct search
for the mirror quarks at the LHC.

%%%%%%%%%%%%%%%%%%%%%%%%%%%%%%%%

\subsubsection{The LHC physics}
%-----------------
\begin{figure}[!t]
	\begin{center}
		{\epsfig{figure=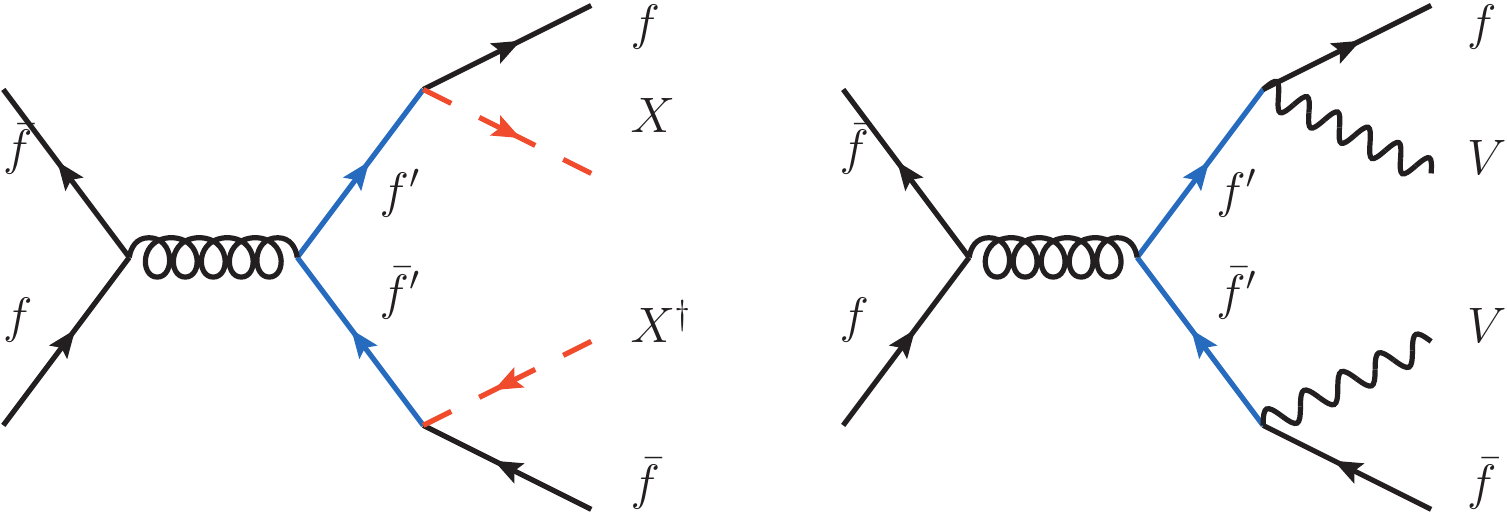,width=0.8\textwidth,height=0.25\textwidth}}
		\caption{Typical Feynman diagrams that produce mirror fermions $f'$ which subsequently decay into SM fermion $f$ and dark matter $X$ or SM gauge boson $V$. \label{fig;LHC}}
	\end{center}
\end{figure}
%------------------------------

In the collider experiments, mirror fermions could be produced if they are light enough. For example, mirror quarks can be pair produced at the LHC. Typical Feynman diagrams are shown in Fig.~\ref{fig;LHC}.  
 The produced mirror fermions decay into a SM fermion, together with DM and/or SM gauge bosons. 
In the case (I) where $X_l$ gets a nonzero VEV, 
the mirror electron decays as $e'\rightarrow l+Z/W/h$  
induced by the mixing with SM leptons. 
The Yukawa coupling $\lambda_u$ induces a decay of the mirror up quark: 
$u'\rightarrow u +X_b$. 

The mirror electron is expected to be the lightest mirror fermion as the SM fermions. 
The type of the daughter lepton depends on the Yukawa couplings $\lambda_e^{ie'}$ 
and that of a daughter boson depends on how it mixes with the SM leptons. 
There are studies about the limits on such extra leptons 
decaying to a lepton and a SM boson~\cite{CMS:2018cgi, 
Falkowski:2013jya,Ellis:2014dza,Dermisek:2014qca,Kumar:2015tna}. 
A conservative limit may be obtained by assuming that the daughter lepton is exclusively a tau lepton. 
In Ref.~\cite{Kumar:2015tna}, 
it is shown that the limit is at most 250 GeV 
with an integrated luminosity about $100\ \text{fb}^{-1}$ at $\sqrt{s} =$ 13 TeV.  
The limit becomes the tightest if a mirror electron exclusively decays to a $Z$-boson 
(and a tau lepton), 
while there is no limit with the integrated luminosity of $100\ \text{fb}^{-1}$
if the mirror electron decays to all of the bosons with a certain branching fraction. 
Therefore the mirror electron above 200 GeV could be allowed by the current data 
at the LHC, although the detail depends on parameters 
which do not have significant correlations with the DM and flavor physics discussed above. 

The mirror up quark should be degenerate with the DM particle in order to explain the relic density, 
and it should dominantly couples to a charm quark (case B) or a top quark (case C). 
In the case~(C), 
since the mass difference is smaller than the W-boson mass, 
the mirror up quark could decay via the four-body decay process, 
\begin{align}
 u' \to t^* + X_b \to W^* + b + X_b \to f_1 f_2 + b + X_b, 
\end{align}
where $W^*$, $t^*$ are off-shell W-boson and top quark, respectively. 
$f_{1,2}$ are the SM fermions coming from $W^*$.
The partial decay width may be so suppressed that 
the two-body decay $u' \to c +X_b$ dominates the mirror up quark decay
even if the Yukawa couplings possess the hierarchy $\lambda_u^{cu'} \ll \lambda_u^{tu'}$. 
Thus the mirror up quark is expected to dominantly decay into a charm quark and a singlet scalar 
in both of the case~(B) and (C).   
The current limit on pair produced top squark decaying to 
a charm quarks and the lightest (neutral) supersymmetric particle is about 500 GeV 
when the mass difference between a top squark and an invisible particle 
is larger than 40 GeV~\cite{Aaboud:2018zjf,Sirunyan:2017kiw}.
The cross section of the mirror up quark pair production is roughly four times larger than  
that of the top squark pair production, 
and a top squark with about 630 GeV gives quarter of a pair production cross section 
of top squark with 500 GeV~\cite{Borschensky:2014cia}. 
Therefore the current limit on the mirror up quark is estimated at about 600 GeV.  

In the case (A), although this case cannot explain the relic density, 
the mirror up quark decays as $u'\rightarrow u +X_b$. 
The signature at the LHC is similar to a squark decaying into an invisible particle and a SM quark which give signals with two jets and missing energy. 
In the mass degenerate region, 
limits on the squark mass is about 650 GeV~\cite{Aaboud:2017vwy,Sirunyan:2018vjp}, 
under the assumption that light-flavor squarks have a common mass. 
The cross section of the up quark pair production is about half of that of the squarks, 
so that the limit is estimated as about 600 GeV~\cite{Borschensky:2014cia}. 
Note that this search is also relevant to the case (B) and the case (C),  
because the difference is if the c-tagging is exploited or not. 
Thus, in any case, we expect that the LHC limits on mirror up quark is about 600 GeV 
in the mass degenerate region. 

The mirror down quark is very long-lived in our model, 
because it does not couple to the SM particles through the Yukawa couplings 
in contrast to the mirror up quark and electron. 
The mirror down quark decays only through the $W_R$-boson exchange, 
so that the decay rate is suppressed by the parity breaking scale $\sim v_R$. 
The decay width is estimated as 
\begin{align}
 \Gamma_{d'} \sim 5.0\times 10^{-22}\  [\text{GeV}] 
\times \left(\frac{v_R}{10^7\ \text{[GeV]}}\right), 
\end{align}  
where the RG corrections to the gauge and Yukawa couplings are neglected. 
The decay width is about three orders of magnitude smaller 
than the decay width of the muon. 

The mirror down quark is expected to be hadronized before it decays, 
and pass through the detector at the LHC. 
This kind of signal is studied in the analyses to search for the so-called R-hadrons which are composite colorless states involving supersymmetric particles~\cite{Aaboud:2016uth,Aaboud:2018hdl}. 
The result in Ref.~\cite{Aaboud:2016uth} gives a limit on the mass of 
bottom squark about 800 GeV, based on a model 
where the R-hadrons are originated from the bottom squark production. 
Since the pair production cross section of the mirror down quark 
is expected to be twice as that of the bottom squark if they have the same masses, 
the limit for the mirror down quark is estimated as about 890 GeV. 
The mirror down quark mass is about eight times heavier than the mirror electron mass 
as expected from the SM fermion masses.  
The current limit may be satisfied even if the mirror electron is about 200 GeV 
and the mirror down quark is about 1.6 TeV. 
This signal is an interesting possibility to be discovered at the LHC in the future.

\subsection{Case (II) : leptonic DM ($X_l$) scenario}\label{sec;XlDM}
Similarly, we can discuss the leptonic DM case where $X_b$ develops a non-vanishing VEV
and $X_l$ does not. In this case, the $Z^R_2$ symmetry remains 
as the remnant of the subgroup of $U(1)^R_{em}$  
and this makes $X_l$ stable. 
Then, $X_l$ is a candidate for cold DM that couples to the charged leptons via
the $\lambda_e$ couplings. In the same manner as Sec.~\ref{sec;XbDM}, we study the DM physics, assuming only the mirror electron makes sizable effects on DM phenomenology because of its light mass, and one of the $\lambda_e$ couplings dominates over the others, e.g., 
$|\lambda_e^{\tau e'}| \gg |\lambda_e^{\mu e'}| , |\lambda_e^{ee'}|$.

\subsubsection{Dark Matter Physics}

%-------------------------------
\begin{figure}[!t]
	\begin{center}
		{\epsfig{figure=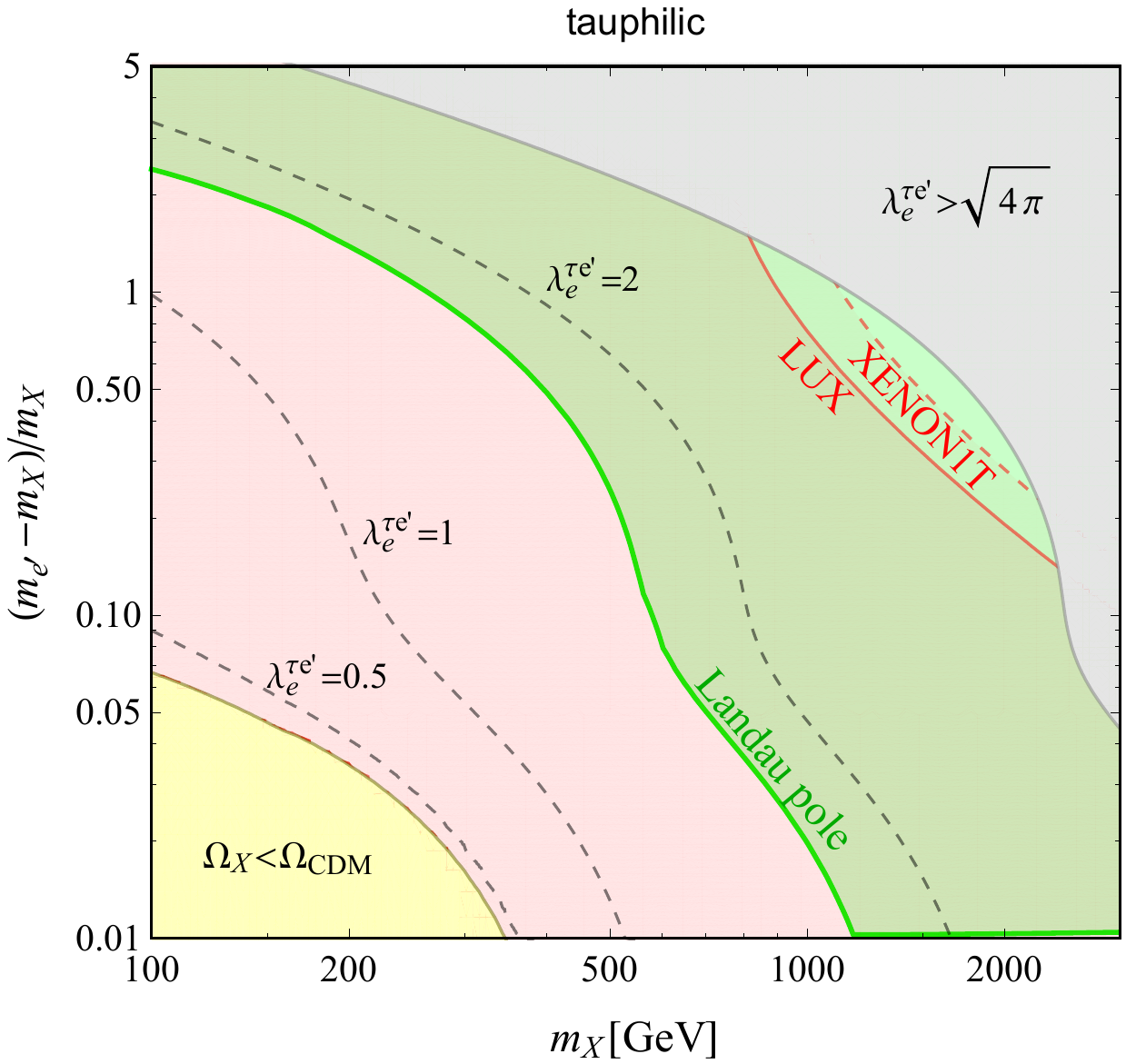,width=0.45\textwidth,height=0.42\textwidth}}{\epsfig{figure=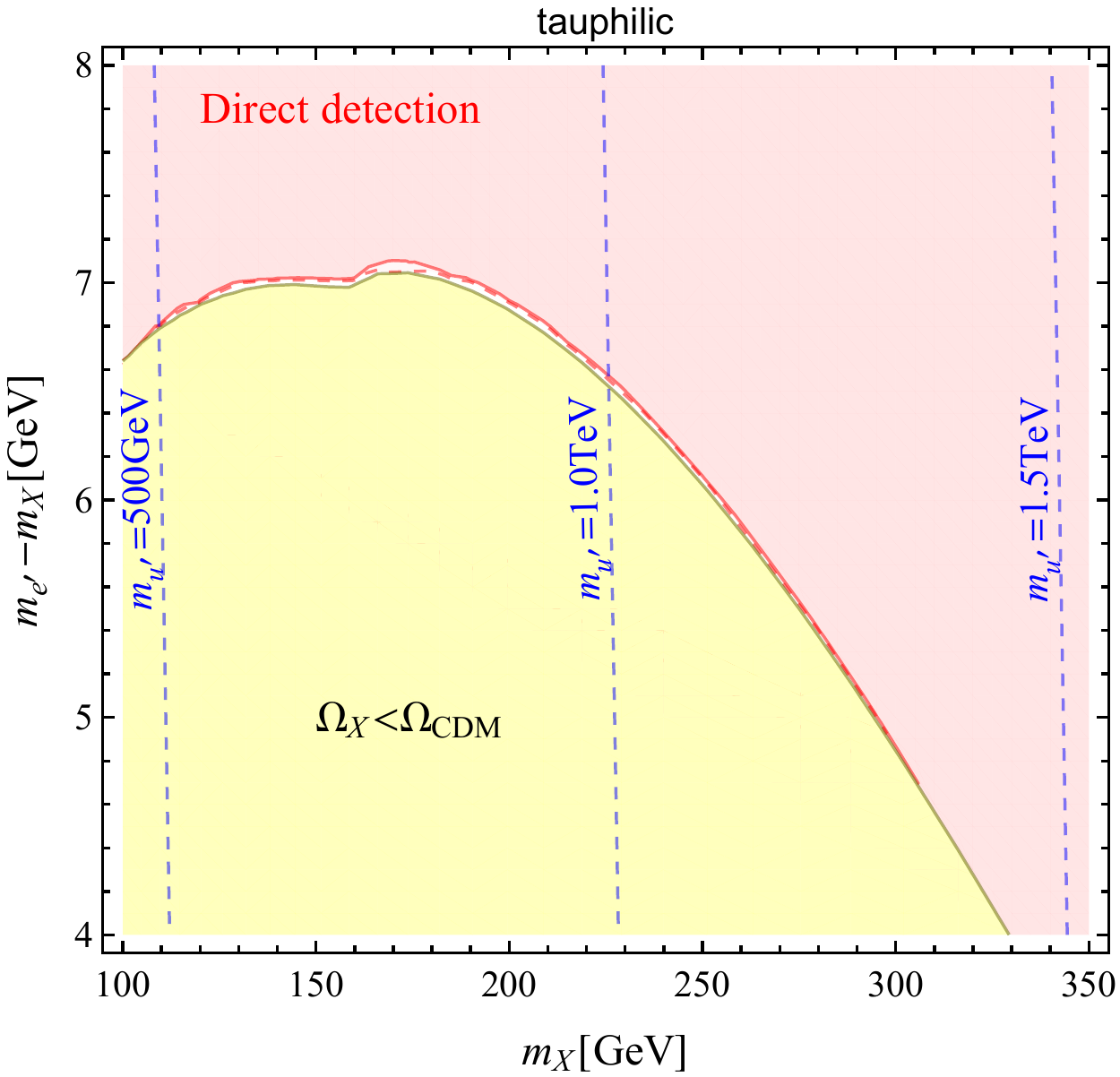,width=0.45\textwidth,height=0.41\textwidth}}
		\caption{$m_X$ vs. the mass difference between $e^\prime$ and $X_l$ in the tau-philic case. Similar to the charm and top-philic cases, we show various contours and the relevant constraints. In the right panel, blue dashed lines show contours for mirror up-quark mass.}
		\label{fig;tau}
	\end{center}
\end{figure}
%------------------------------

The DM physics in the leptophilic scenario can be understood directly from analysis in the case (I). 
The annihilation is $p$-wave dominant due to the light charged lepton masses, 
so that Yukawa couplings have to be large enough to account for the DM abundance. 
Direct direction is simple as well. The DM-nuclei scattering is caused only through photon and $Z$ exchanging, 
because the DM particle does not directly couple to any colored particles. 
Besides, the light lepton masses lead the negligible $Z$-exchanging contribution. 
Thus, the main process is the photon-exchanging in the whole parameter space. 

We would like to note, however, 
that we need a modification in Eq.~(\ref{eq:photon_exchange}) 
in the electron-philic case.   
This equation was derived assuming the momentum transfer is negligibly small 
compared to particle masses in the loop. 
The typical transferred momentum is $\sim 50$\,MeV for xenon detectors, for example. 
Therefore, Eq.~(\ref{eq:photon_exchange}) is invalid in the electron-philic case, 
and we have to modify the expression by taking a finite momentum transfer into account.

Figure~\ref{fig;tau} shows how parameter space is constrained in the tau-philic case in the same manner as in the charm-philic case. There is no big difference between the muon-philic and the tau-philic case. 
The electron-philic case is strongly constrained by the EW precision measurement, given by the LEP experiment.
In the white region, the relic density is explained without any conflicts with all the constraints, 
but it is very narrow.

\subsubsection{Flavor Physics}
In the leptonic DM case, we assume that one element of $\lambda^{ij}_e$ is sizable.
In such a case,
one of the stringent constraints comes from $l_2 \to l_1 \, \gamma$.
The processes are given by the dipole operators:
\beq
{\cal L}_e = - C^{ij}_{e } \left [ \frac{e}{16 \pi^2} \, m^j_{e} \, \overline{e^i_R} \, \sigma_{\mu \nu} e^j_L  \, F^{\mu \nu} \right ]. 
\eeq
$  C^{ij}_{e }$ is estimated at the one-loop level following the result on the exotic top decay in Sec. \ref{sec;flavor1}: 
\beq
  C^{ij}_{e }=- \frac{\lambda^{ik}_e \lambda^{jk \, *}_e }{24  \, m_X^2}\, f_7 (x_k).
\eeq
In particular, the flavor-violating muon decay, i.e. $\mu \to e \, \gamma$, is severely constrained
by the experiment: $Br(\mu \to e \, \gamma)< 4.2\times10^{-13}$ \cite{TheMEG:2016wtm}.
In the muon-philic DM or electron-philic DM case, the exotic decay 
may be enhanced by the Yukawa coupling.
The upper bound on $\lambda^{ij}_e$ is estimated as
\beq
\left |\lambda^{e e'}_e \lambda^{\mu e'}_e \right | \lesssim 0.002~(0.009),
\eeq
when $m_{e^\prime} \approx m_X$ is imposed and $m_X$ is fixed at 500 GeV (1 TeV).

The flavor-violating $\tau$ decays, i.e. $\tau \to e \, \gamma$, $\mu \, \gamma$, are
less constrained and we confirm that our predictions are below the current bounds
which are ${\cal O}(10^{-8})$ \cite{PDG} even if the Yukawa couplings are ${\cal O}(1)$.

As another process, 
the lepton flavor violating (LFV) process, $l^j \to l^i \, \overline{l^k} \, l^{k^\prime} $ may become sizable depending on setups. The contribution to the process is given by two types of diagrams: the box diagram and
the penguin diagram. The box diagram is much suppressed in our setup,
since it is linear to one sizable $\lambda^{j e'}_e$ and suppressed three couplings, $\lambda^{i e'}_e \lambda^{k e'}_e \lambda^{k^\prime e'}_e$, for instance.
If  $l^{k^\prime}$ is identical to $l^{k}$ or $l^{i}$, the penguin diagram is possible.
We can estimate the upper bound on the Yukawa coupling, using the experimental upper bound on the muon decay, $\mu \to 3 e$ as~\cite{Bellgardt:1987du,Perrevoort:2016nuv} 
\beq
\left |\lambda^{e e'}_e \lambda^{\mu e'}_e \right | \lesssim 0.06~(0.23),
\eeq
when $m_{e^\prime} \approx m_X$ is imposed and $m_X$ is fixed at 500 GeV (1 TeV).
Thus, we conclude that the bound from $\mu \to e \gamma$ is more important in our model.
Similarly, we can discuss the LFV decays of $\tau$, but our prediction is 
much below the experimental bound, even if the Yukawa couplings are assumed to be ${\cal O}(1)$.
We have also estimated our prediction on the $\mu$-e conversion process in nuclei,
but the bound is also not stronger than the one from $\mu \to e \gamma$.

\subsubsection{The LHC physics}
The relevant processes for the mirror fermions are also given by Fig.~\ref{fig;LHC} as in the case (I), 
while the mirror electron decays as $e'\rightarrow l+X_l$ 
and  the mirror up quark decays as $u'\rightarrow u +V$. 
The signal of the mirror down quark is not changed from the case (I). 

In this case, the mirror electron pair production gives the same signal as the slepton pair production. 
In parameter space where a mass difference between a slepton and an invisible particle is less than 10 GeV, 
the lower limit is at most 190 GeV under the assumption that 
all of the selectron and smuon have a common mass~\cite{Aaboud:2017leg}. 
This limit is expected to be directly applicable to the mirror electron, 
because the production cross section is similar 
to the sleptons pair production in the analysis~\cite{Aaboud:2017leg}. 
There may be no limit in the moderate mass difference region.  
If the mass difference is about 50 GeV (200 GeV), the lower limit is about 200 (500) GeV 
under an assumption of degenerate selectron, smuon and stau~\cite{Aaboud:2018jiw,Sirunyan:2018nwe}.
The limits will be slightly weaker for the mirror electron in this model due to the smaller production cross section. 
If the mirror electron exclusively decays to a tau lepton and a DM particle, 
the signal will be too small to give bounds on the mirror electron mass~\cite{Sirunyan:2018vig}. 

The signature of the mirror quark decay $u' \to q + Z/W/h$ will be similar 
to the vector-like quark searches~\cite{Aaboud:2018pii,CMS:2018haz}. 
As for the $e' \to l + Z/W/h$ in the case (I), the limit depends on flavor of a daughter quark 
and a type of daughter boson.
The limits on vector-like partners of the third-generation quarks are studied in Ref.~\cite{Aaboud:2018pii}, 
and a limit on a top (bottom) quark partner is 1.31 (1.03) TeV for any combination of decay modes.

\section{Neutrino sector}   
\label{sec;neutrino}

%%%%%%%%%%%%%%%%%%%%%%%%%%%%%%%%%%%%%%%%%%%%%%%%%%%%%%%%%
\begin{table}[h]
	\begin{center}
		\begin{tabular}{ccccccc}
			\hline
			\hline
			Fields & spin   & ~~$SU(3)_c$~~ & ~~$SU(2)_L$~~& ~~$SU(2)_R$~~  & ~~$U(1)_R$  & ~~$U(1)_L$      \\ \hline  
			$N^{ i} $ & $1/2$ &  ${\bf 1}$   &${\bf1}$       &${\bf1}$      &         $0$   &       $0$      \\ 
			\hline 
		\end{tabular}
	\end{center}
	\caption{Neutrinos in the model. $i$ denotes the flavors: $i=1, \, 2, \, 3$. }
	\label{table4}
\end{table} 
%%%%%%%%%%%%%%%%%%%%%%%%%%%%%%%%%%%%%%%%%%%%%%%%%%%%%%%%%
In this section, we discuss the neutrino sector. Since neutrino oscillation experiments indicate that at least two of three SM neutrinos have to be massive, we need accommodate massive neutrinos in this model. To accomplish that, we introduce fermionic fields $N^i$ that are neutral under the gauge symmetry as shown in Table \ref{table4}.
$N^i$ are transformed by the parity as
\beq
P \, N^i(t, x) \, P =  \gamma_0 \, N^i (t,-x).
\eeq
Then, the Yukawa couplings concerned with the neutrino masses are given by
\beq
Y^{ij} \left ( \overline{l^i_L} \widetilde H_L N^j_R + \overline{l^{\prime \, i}_R} \widetilde H_R N^j_L \right ) +h.c.. 
\eeq
In addition, we can write down all the possible mass terms:
\beq
m^{ij} \overline{N^{ i}} N^{ j} + \frac{1}{2} \, M^{ij} \overline{N^{ i \, c}} N^{ j} +h.c..
\eeq
The first and second terms correspond to the Dirac and Majorana mass terms, respectively. 
Depending on the size of each mass term, we can discuss some possibilities.
If $M^{ij} $ vanishes, the active neutrinos are  Dirac fermions.
If $m^{ij} $ vanishes, the active neutrinos are  Majorana fermions.
We consider both cases below.

\subsection{Dirac neutrino scenario}
If $N^i$ is expected to be charged under the U(1) lepton symmetry,
the Majorana mass matrix, $M^{ij}$, is forbidden.
Assuming $|m^{ij}| \gg v_R$, we obtain the tiny Dirac mass matrix for the active neutrinos:
\beq
\left ( Y m^{-1} Y^{\dagger} \right )^{ij} \left ( \overline{l^i_L} \widetilde H_L   H_R \sigma_2 l^{\prime \, j}_R  \right ) +h.c.. 
\eeq
Thus, the active neutrino mass matrix, $m_\nu$, is evaluated as
\beq
(m_\nu)^{ij}=  \frac{v_L \,v_R}{2} \left ( Y m^{-1} Y^{\dagger} \right )^{ij}.
\eeq

Naively, introducing many extra neutrino states would be in conflict with current cosmological bounds on neutrino masses and the number of light species. However, we can actually show that all those new components were never produced abundantly in the early Universe if the relevant couplings are small or the new mass scales are high. In the Dirac neutrino case, the three right-handed neutrinos $\nu'_R$ could be abundantly produced due to its $SU(2)_R$ gauge interaction if the universe was hot enough, but they will decouple earlier. They will contribute to the effective number of neutrinos by
\begin{equation}\label{eq:neff}
\delta  N_{\textrm{eff}} = 3\times \frac{T^4_{\nu'_R}}{T^4_{\nu_L}}=3\left[ \frac{g_{\ast s}\left(T_{\nu_L}\right)}{g_{\ast s}\left(T_{\textrm{dec}}\right)}
\right]^{4/3},
\end{equation}
where $T_{\nu_L}$ is the three active left-handed neutrino's temperature, equal to $\left(4/11\right)^{1/3}T_\gamma$ after $e^\pm$'s annihilation, $T_{\textrm{dec}}$ for the temperature at which $\nu'_R$ decouples from the SM sector, $g_{\ast s}$ counts the effective number of degrees of freedom for entropy density in the SM. We can get a lower bound on $\delta  N_{\textrm{eff}}$ before $e^\pm$'s annihilation when the decoupling temperature $T_{\textrm{dec}}$ is larger than top quark's mass,
\begin{equation}\label{eq:neff2}
\delta  N_{\textrm{eff}} \simeq 3\left[ \frac{43/4}{427/4}\right]^{4/3}= 0.14,
\end{equation} 
This number is well below the Planck's limit, $\delta  N_{\textrm{eff}}\lesssim 0.30$~\cite{Planck:2018eyx}. Since the $\nu'_R$ and $\nu_L$ combine into a Dirac neutrino, they would have the same mass. Therefore, current cosmological limit on neutrino's mass is also safe in this scenario. Future experiments would have the sensitivity to probe $\delta N_{\textrm{eff}} \sim 0.02-0.03$~\cite{Abazajian:2016yjj}. 

\subsection{Majorana neutrino scenario}
If the lepton symmetry is not assigned, the Majorana mass matrix is allowed
and the tiny neutrino masses can be generated by integrating out the heavy $N^i$:
\beq
\left ( Y M^{-1} Y^{T} \right )^{ij} \left ( \overline{l^i_L} \widetilde H_L   H_L \sigma_2 l^{c \, j}_L+ \overline{l^{\prime \, i}_R} \widetilde H_R   H_R \sigma_2 l^{\prime c \, j}_R  \right ) +h.c.,
\eeq
in the limit that the Dirac mass matrix, $m^{ij}$, is vanishing.
The first term gives the active neutrino masses.

Cosmological constraint in this case becomes more complicated than that in the Dirac case. 
We have two copies of seesaw mechanism. 
Then for each generation, after mass diagonalization we would have one active Majorana neutrino and three new Majorana ones. Two of the three new neutrinos can be very heavy due to the Majorana mass $M_{ij}$ and they can decay quickly. The remaining one has a mass $m_\nu\times v^2_R/v^2_L$ where $0\leq m_\nu\lesssim 0.1$ eV. Since we expect $v_R/v_L>10^6$ at least, we would have a $\mathcal{O}(100)$ GeV neutrino for $m_\nu\sim 0.1$ eV. If abundant and stable, they will contribute too much to the energy density and overclose our Universe. Fortunately, one of the three active neutrinos can be massless, which also means one of the three mirror neutrinos can be massless. Since neutrino mixing occurs also in the mirror sector, we would have the decay process for heavy mirror neutrinos ($\nu'_H$) into three massless mirror neutrinos ($\nu'_0$) through the $Z_R$ mediator, $\nu'_H \rightarrow 3 \nu'_0$, with decay width
\begin{equation}\label{eq:nudecay}
\Gamma \sim \frac{m^5_{\nu'_H}}{32\pi m^4_{Z_R}}\simeq 2.2 \textrm{ s}^{-1}\times \left(\frac{v_R/v_L}{10^6}\right)^6\left(\frac{m_\nu}{0.1\textrm{eV}}\right)^5.
\end{equation}
As long as $\Gamma \gtrsim 1 \textrm{ s}^{-1}$, the heavy mirror neutrinos will decay before BBN era. However, $\nu'_H$ should decay where it is still relativistic, otherwise the decay products would constitute too much dark radiation. This put a constraint on 
\begin{equation}
\Gamma \gtrsim 10^{10}\textrm{ s}^{-1} \left(\frac{v_R/v_L}{10^6}\right)^4\left(\frac{m_\nu}{0.1\textrm{eV}}\right)^2.
\end{equation}
Combined with Eq.(\ref{eq:nudecay}), it would give $v_R/v_L> 10^{11}$, a very stringent limit. Note that if the reheating temperature after inflation is low, those mirror particles might not be produced abundantly, since these heavy mirror neutrinos can be in thermal equilibrium only above the temperature, 
$\sim  (v_R/v_L \cdot 10^{-6})^{4/3}\times 100\ \textrm{TeV}$. In such a case, the above bound could be much relaxed. 

\section{Summary}
\label{sec;summary}

We have proposed an extended Standard Model (SM) with parity symmetry, motivated 
by the strong CP problem and DM in our universe. 
The SM gauge symmetry is enlarged at high energy scale to 
$SU(3)_c \times SU(2)_L \times U(1)_L \times SU(2)_R \times U(1)_R$. 
The mirror quarks, leptons and Higgs bosons are introduced, 
and they belong to the same representations 
of the mirror EW gauge symmetry $SU(2)_R\times U(1)_R$ 
as in the SM.  
The model respects parity symmetry at high energy scale 
where the mirror sector is not decoupled. 
In addition to the minimal extension, 
two scalar fields, $X_b$ and $X_l$, are added to the model. 
These scalar fields provide portals between the SM and mirror sector, 
and resolve the abundance problem of the lightest stable mirror quark and lepton.   
Interestingly, either of the scalar fields is a good candidate for DM, 
since a remnant of the gauge symmetry $U(1)_L\times U(1)_R$ 
guarantees its stability.

We have investigated various phenomenologies in this model, 
including direct detection of the DM, flavor physics, collider physics 
and cosmological effects. 
One special prediction of this model is the mass ratio of mirror particles, which might be probed at future collider searches. At this moment, the LHC data has already constrained the mass of the lightest mirror quark, $\gtrsim \mathcal{O}(600)$GeV. 
This indicates that the parity breaking scale should be larger than $7\times 10^{7}$\,GeV. 
In addition, the mirror up quark has to be lighter than about 3\,TeV 
in order to avoid the strong bound from the direct detection for the DM,  
if the observed DM abundance is saturated with the scalar DM in our model. 
We showed that the mass splitting of the DM 
and the mirror up quark or electron is required to be tuned at ${\cal O}$(1-10\,\%) level.  
This conclusion is not changed, 
even if the contribution of the Higgs portal coupling is taken into account. 
There is an upper bound on the parity symmetry breaking scale, 
$\lesssim4\times10^8$\,GeV 
which comes from the upper bound on the mirror up quark mass to explain the DM.  
The mirror electron resides around 200-400 GeV in this parameter region. 
Such a mass region requires more detailed analyses 
of the flavor physics, collider physics and the EW precision measurement.  
The right-handed neutrinos contribute to the effective relativistic degree of freedom 
so large that it can be tested by the future CMB experiments.

\section*{Acknowledgment}
This work is supported in part by the Grant-in-Aid for Scientific
Research from the Ministry of Education, Science, Sports, and Culture (MEXT), Japan (No.18K13534 [JK] and No.17H05404 [YO]), 
the Department of Energy under Award No.~DE-SC0011276~[JK] of the U.S., 
JSPS Overseas Challenge Program for Young Researchers
and NSERC of Canada [SO], and the Grant-in-Aid for Innovative Areas, Japan (No.16H06490 [YT]).

\appendix

 \section{RG equations below the parity breaking scale}
 \label{appendixA}
In this appendix, we summarize the RG equations for the relevant couplings in our model.

The beta-functions for the gauge couplings are 
\begin{align}
 16\pi^2 \beta_{g_Y} =&\ g_Y^3 \left[  \frac{41}{6} 
  + \frac{4}{3} \sum_{i=1}^3  
    \left( \frac{4}{3} \theta_{u'_i}+\frac{1}{3}\theta_{d'_i} + \theta_{e'_i} \right)   
  \right], \\
 16\pi^2 \beta_{g_{2}} =&\ -\frac{19}{6} g_2^3, \\
  16\pi^2 \beta_{g_3} =&\ g_3^3 \left[ 
  -7 + \frac{2}{3}\sum_i \left(\theta_{u'_i}+\theta_{d'_i}\right)
   \right], 
\end{align}
where $\theta_{\phi} = 1$ for $\mu > m_\phi$ and $\theta_\phi=0$ for $\mu < m_\phi$. 

The beta functions for the Yukawa couplings are given by 
\begin{align}
 16\pi^2 \beta_{Y_u} =&\ \frac{3}{2} \left(Y_u Y_u^\dagger - Y_d Y_d^\dagger \right) Y_u 
                                    + \frac{1}{2} Y_u \lambda_u^\dagger \lambda_u 
                                    + Y_2(H) Y_u - 
                                    \left( \frac{17}{12} g_Y^2 + \frac{9}{4} g^2 + 8 g_s^2
                                   \right) Y_u, \\ 
 16\pi^2 \beta_{Y_d} =&\ \frac{3}{2} \left(Y_d Y_d^\dagger - Y_u Y_u^\dagger \right) Y_d 
                                    + Y_2(H) Y_d - 
                                    \left( \frac{5}{12} g_Y^2 + \frac{9}{4} g^2 + 8 g_s^2
                                   \right) Y_d, \\ 
 16\pi^2 \beta_{Y_e} =&\ \frac{3}{2} Y_e Y_e^\dagger Y_e 
                                    + \frac{1}{2} Y_e \lambda_e^\dagger \lambda_e  
                                    + Y_2(H) Y_e - 
                                    \left( \frac{15}{4} g_Y^2 + \frac{9}{4} g^2 
                                   \right) Y_e, \\ 
 16\pi^2 \beta_{\lambda_u} =&\  \lambda_u \left( 
                                     \lambda_u^\dagger \lambda_u + Y_u^\dagger Y_u \right) 
                                    + Y_2(X_b) \lambda_u - 
                                    \left( \frac{8}{3} g_Y^2 + 8 g_s^2
                                   \right) \lambda_u, \\ 
 16\pi^2 \beta_{\lambda_e} =&\  \lambda_e \left( 
                                     \lambda_e^\dagger \lambda_e + Y_e^\dagger Y_e \right) 
                                    + Y_2(X_l) \lambda_e - 
                                    6 g_Y^2  \lambda_e,  
\end{align}
where 
\begin{align}
 Y_2(H) =&\ \text{Tr} \left( 3Y_u^\dagger Y_u + 3Y_d^\dagger Y_d 
                                             + Y_e^\dagger Y_e \right), \\
 Y_2(X_b) = &\ \text{Tr} \left( 3\lambda_u^\dagger \lambda_u \right), \quad  
 Y_2(X_l) =  \text{Tr} \left( \lambda_e^\dagger \lambda_e \right) 
\end{align}
are defined. 
The decoupling effects can be included by replacing 
\begin{align}
 \lambda_{u}^{ij} \to  \lambda_{u}^{ij} \theta_{u'_i} \theta_{X_b}, \quad 
 \lambda_{e}^{ij} \to  \lambda_{e}^{ij} \theta_{e'_i} \theta_{X_l}. 
 \end{align}

In Figs.~\ref{fig;charm}, \ref{fig;top} and \ref{fig;tau}, 
the perturbativity bounds are calculated based on these RGEs. 
We solved the 2-loop SM RGE up to the DM mass scale, 
then the relevant contributions of the mirror fermions 
are added to the beta functions step by step. 
The mirror fermion masses are determined by the Yukawa couplings at the parity breaking scale.
Hence, the RG running changes the mirror fermion masses themselves 
which determine scales where the contributions are turned on. 
We solve the RG equations iteratively until 
\begin{align}
 \sum_{f'} \left|\frac{m_{f'}^{(N)} - m_{f'}^{(N-1)}}{m_{f'}^{(N)}}\right| < 10^{-4}
\end{align}
is satisfied, where $N$ is the number of loops of the numerical calculation
 and $m^{(N)}_{f'}$ is a mirror fermion mass in the $N$-th loop. 
We define a parameter point as non-perturbative 
if any coupling blows-up during this iterative procedure
or any coupling is larger than $\sqrt{4\pi}$ at any scale.

\section{Role of Higgs portal interaction}
\label{appendixB}

In this model, the DM candidate, $X$, can have a non vanishing Higgs portal coupling. 
Here, we shall clarify how much this coupling improves the constraints. 

For Higgs portal process, annihilation is $s$-wave dominant, 
while for mirror fermion exchanging it is $p$-wave dominant. 
Since there is no interference between different partial waves, 
the annihilation cross section can be separated into two contributions\footnote{For $m_X \gtrsim 1$\,TeV, annihilation via Higgs portal is dominated by $XX^\dagger \to WW,ZZ,hh$ processes, so that there is no large interference in $XX^\dagger\to t\bar{t}$ process. Hence, in such a mass region, once we replace like $b_f v^2 \to (a_f + b_f v^2)$, discussion below can be applied straightforwardly even in top-philic case.}, 
\begin{equation}
\sigma v \simeq a_h \lambda_{hX}^2 + b_f v^2 |\lambda|^4 ,
\end{equation}
where $\lambda_{hX}$ and $\lambda$ are a Higgs portal coupling and a $X$-$\psi$-$\psi'$ Yukawa coupling  given in Eq.(\ref{eq;lambda}), respectively.
This is valid except for coannihilation region. 
Suppose that the Higgs portal contribution is $r$ times larger than that of mirror fermion exchanging 
at freeze-out, i.e., 
\begin{equation}
a_h \lambda_{hX}^2 = r b_f v^2 |\lambda|^4 ,
\label{eq;ratio}
\end{equation}
the cross section is rewritten as 
\begin{equation}
\sigma v \simeq (1+r) b_f v^2 |\lambda|^4 .
\end{equation}
To explain the DM abundance, $\lambda$ should satisfy the equation, 
\begin{equation}
 (1+r) b_f v^2 |\lambda|^4  \simeq 10^{-9} \, [{\rm GeV^{-2}]}.
\label{eq;def_lamr}
\end{equation}
We define $\lambda$ satisfying Eq.(\ref{eq;def_lamr}) as $\lambda(r)$. 
It is easy to see that $\lambda(r)$ is related to $\lambda(0)$ as 
\begin{equation}
\lambda (r) = \frac{ \lambda(0) }{ (1+r)^{1/4} } .
\label{eq;lamr}
\end{equation}
This means for a nonzero Higgs portal coupling, $\lambda$ required to 
explain the DM abundance is smaller by a factor, $1/(1+r)^{1/4}$, than the vanishing case. 

The SI cross section of WIMP DM and nuclei scattering is given by 
\begin{equation}
\sigma_{SI} = \frac{\mu^2}{\pi} [ Z f_{p,e} + (A-Z) f_{n,e} ]^2 
+ \frac{\mu^2}{\pi} [ Z f_{p,o} + (A-Z) f_{n,o} ]^2 ,
\end{equation}
where $\mu = m_A m_X / (m_A + m_X)$ denotes reduced mass of DM and nucleus and 
we assume the symmetric relic, i.e. $\Omega_{DM} = \Omega_{\overline{DM}}$. 
$f_{N,e}$ contains only $C_S$ and $C_T$ and $f_{N,o}$ does only $C_V$ in Eq.(\ref{eq;LeffDD}). 
Note that the $\lambda$ coupling mainly generates $f_{N,o}$ 
through photon and $Z$ boson exchanging, while the Higgs portal interaction only $f_{N,e}$. 
Then, in our model, it takes a form of
\begin{equation}
\sigma_{SI}(r) = C_{HP} \lambda_{hX}^2 + C_{FP} |\lambda(r)|^4 .
\end{equation}
From this equation, if $\sigma_{SI}(r) < \sigma_{SI}(0) = C_{FP} |\lambda(0)|^4$, 
we conclude the Higgs portal interaction can relax the direct detection constraints. 
Using Eqs.(\ref{eq;ratio}), (\ref{eq;def_lamr}) and (\ref{eq;lamr}), we obtain 
\begin{align}
\sigma_{SI}(r) 
& = \frac{r}{1+r} \sigma_{HP} + \frac{1}{1+r} \sigma_{SI}(0) ,
\end{align}
where 
\begin{equation}
\sigma_{HP} \simeq C_{HP} \times \frac{10^{-9}\,[{\rm GeV}^{-2}]}{a_h} 
\end{equation}
corresponds to the SI cross section for the pure Higgs portal DM scenario and is estimated as $\sigma_{HP} \simeq 1.7 \times 10^{-9}$\,[pb] when $m_X \gtrsim {\cal O}$(TeV). 
Thus, the SI cross section is reduced only if 
\begin{equation}
a \equiv \frac{\sigma_{HP}}{\sigma_{SI}(0)} < 1.
\label{eq;condition}
\end{equation}
The reduction rate is evaluated as
\begin{equation}
\frac{\sigma_{SI}(r)}{\sigma_{SI}(0)} = \frac{1+a \, r}{1+r}.
\end{equation}
We would like to point out that the modified cross section is bounded:
\begin{equation}
\sigma_{HP} \leq \sigma_{SI} \leq \sigma_{SI}(0).
\end{equation}
This indicates that a mass region already excluded in the Higgs portal scenario is 
not rescued even if we introduce Higgs portal interaction. 
The current XENON1T result rules out the DM mass $\lesssim 2$\,TeV for complex scalar DM in the Higgs portal scenario. 

For example, in charm-philic case, $\sigma_{SI}(0)$ is 
\begin{equation}
\sigma_{SI}(0) \simeq 1.1 \times 10^{-9} \, [{\rm pb}] ,
\end{equation}
for $m_X=3$\,TeV and $m_{u'}=4$\,TeV. 
This is comparable with the Higgs portal one, and hence 
direct detection bound is not relaxed in this parameter region. 
The value of $\lambda$ is reduced to, e.g.
\begin{equation}
\begin{split}
\lambda(r=4) & \simeq \frac{\lambda(0)}{5^{1/4}} \simeq 2.2 ,\quad
\lambda(r=10) \simeq \frac{\lambda(0)}{11^{1/4}} \simeq 1.8 \,.
\end{split}
\end{equation}
This will still suffer from Landau pole constraint, however. 
Thus, we do not expect improvement in charm-philic case. 

In a similar way, we find 
\begin{equation}
\sigma_{SI}(0) \simeq 2.1 \times 10^{-8} \, [{\rm pb}] ,
\end{equation}
for $m_X=3$\,TeV and $m_{u'}=4$\,TeV in the top-philic case.
This leads to $\sigma_{SI}(r=4) \simeq 5.6 \times 10^{-9}$\,[pb], but it is still above the XENON1T bound. 
When $r=4$, the $\lambda$ value is 
\begin{equation}
\lambda(r=4) \simeq \frac{\lambda(0)}{5^{1/4}} \simeq 2.1 .
\end{equation}
Therefore, in the top-philic case, we can expect some reduction of the SI cross section 
by introducing a nonzero $\lambda_{hX}$, but it is not enough to evade various constraints.

\section*{ }


\begin{thebibliography}{99}
%\cite{Baker:2006ts}
\bibitem{Baker:2006ts} 
  C.~A.~Baker {\it et al.},
  %``An Improved experimental limit on the electric dipole moment of the neutron,''
  Phys.\ Rev.\ Lett.\  {\bf 97}, 131801 (2006)
%  doi:10.1103/PhysRevLett.97.131801
  [hep-ex/0602020].
  %%CITATION = doi:10.1103/PhysRevLett.97.131801;%%
  %1084 citations counted in INSPIRE as of 09 Dec 2018

	
\bibitem{Peccei:1977hh} 
R.~D.~Peccei and H.~R.~Quinn,
%``CP Conservation in the Presence of Instantons,''
Phys.\ Rev.\ Lett.\  {\bf 38}, 1440 (1977).
%doi:10.1103/PhysRevLett.38.1440
%%CITATION = doi:10.1103/PhysRevLett.38.1440;%%

\bibitem{Peccei:1977ur} 
R.~D.~Peccei and H.~R.~Quinn,
%``Constraints Imposed by CP Conservation in the Presence of Instantons,''
Phys.\ Rev.\ D {\bf 16}, 1791 (1977).
%doi:10.1103/PhysRevD.16.1791
%%CITATION = doi:10.1103/PhysRevD.16.1791;%%

\bibitem{Weinberg:1977ma} 
S.~Weinberg,
%``A New Light Boson?,''
Phys.\ Rev.\ Lett.\  {\bf 40}, 223 (1978).
%doi:10.1103/PhysRevLett.40.223
%%CITATION = doi:10.1103/PhysRevLett.40.223;%%

\bibitem{Wilczek:1977pj} 
F.~Wilczek,
%``Problem of Strong  $P$  and  $T$  Invariance in the Presence of Instantons,''
Phys.\ Rev.\ Lett.\  {\bf 40}, 279 (1978).
%doi:10.1103/PhysRevLett.40.279
%%CITATION = doi:10.1103/PhysRevLett.40.279;%%

%\cite{Babu:1989rb}
\bibitem{Babu:1989rb} 
  K.~S.~Babu and R.~N.~Mohapatra,
  %``A Solution to the Strong {CP} Problem Without an Axion,''
  Phys.\ Rev.\ D {\bf 41}, 1286 (1990).
%  doi:10.1103/PhysRevD.41.1286
  %%CITATION = doi:10.1103/PhysRevD.41.1286;%%
  %148 citations counted in INSPIRE as of 29 Jan 2018


%\cite{Barr:1991qx}
\bibitem{Barr:1991qx} 
  S.~M.~Barr, D.~Chang and G.~Senjanovic,
  %``Strong CP problem and parity,''
  Phys.\ Rev.\ Lett.\  {\bf 67}, 2765 (1991).
%  doi:10.1103/PhysRevLett.67.2765
  %%CITATION = doi:10.1103/PhysRevLett.67.2765;%%
  %97 citations counted in INSPIRE as of 09 Mar 2018


%\cite{Gu:2012in}\cite{Abbas:2017vle}\cite{Abbas:2017hzw}\cite{Gu:2017mkm}
\bibitem{Gu:2012in} 
  P.~H.~Gu,
  %``Mirror left-right symmetry,''
  Phys.\ Lett.\ B {\bf 713}, 485 (2012)
 % doi:10.1016/j.physletb.2012.06.042
  [arXiv:1201.3551 [hep-ph]].
  %%CITATION = doi:10.1016/j.physletb.2012.06.042;%%
  %17 citations counted in INSPIRE as of 29 Jan 2018
  
    %\cite{Abbas:2017vle}
\bibitem{Abbas:2017vle} 
  G.~Abbas,
  %``A model of spontaneous $CP$ breaking at low scale,''
  Phys.\ Lett.\ B {\bf 773}, 252 (2017)
%  doi:10.1016/j.physletb.2017.08.028
  [arXiv:1706.02564 [hep-ph]].
  %%CITATION = doi:10.1016/j.physletb.2017.08.028;%%
  %5 citations counted in INSPIRE as of 29 Jan 2018
%\cite{Abbas:2017hzw}
\bibitem{Abbas:2017hzw} 
  G.~Abbas,
  %``A low scale left-right symmetric mirror model,''
  arXiv:1706.01052 [hep-ph].
  %%CITATION = ARXIV:1706.01052;%%
  %5 citations counted in INSPIRE as of 29 Jan 2018

  
  %\cite{Gu:2011yx}
\bibitem{Gu:2011yx} 
  P.~H.~Gu,
  %``A left-right symmetric model with SU(2)-triplet fermions,''
  Phys.\ Rev.\ D {\bf 84}, 097301 (2011)
%  doi:10.1103/PhysRevD.84.097301
  [arXiv:1110.6049 [hep-ph]].
  %%CITATION = doi:10.1103/PhysRevD.84.097301;%%
  %2 citations counted in INSPIRE as of 14 May 2018
  
  %\cite{Gu:2013gic}
\bibitem{Gu:2013gic} 
  P.~H.~Gu,
  %``Mirror symmetry: from active and sterile neutrino masses to baryonic and dark matter asymmetries,''
  Nucl.\ Phys.\ B {\bf 874}, 158 (2013)
 % doi:10.1016/j.nuclphysb.2013.05.013
  [arXiv:1303.6545 [hep-ph]].
  %%CITATION = doi:10.1016/j.nuclphysb.2013.05.013;%%
  %7 citations counted in INSPIRE as of 14 May 2018


  %\cite{Gu:2017mkm}
\bibitem{Gu:2017mkm} 
  P.~H.~Gu,
  %``Spontaneous mirror left-right symmetry breaking for leptogenesis parametrized by Majorana neutrino mass matrix,''
  JHEP {\bf 1710}, 016 (2017)
%  doi:10.1007/JHEP10(2017)016
  [arXiv:1706.07706 [hep-ph]].
  %%CITATION = doi:10.1007/JHEP10(2017)016;%%
  %3 citations counted in INSPIRE as of 29 Jan 2018
  
  
  %\cite{DAgnolo:2015uqq}
\bibitem{DAgnolo:2015uqq} 
  R.~T.~D'Agnolo and A.~Hook,
  %``Finding the Strong CP problem at the LHC,''
  Phys.\ Lett.\ B {\bf 762}, 421 (2016)
  %doi:10.1016/j.physletb.2016.09.061
  [arXiv:1507.00336 [hep-ph]].
  %%CITATION = doi:10.1016/j.physletb.2016.09.061;%%
  %7 citations counted in INSPIRE as of 14 May 2018
 %\cite{Hall:2018let}

%\cite{Alvarado:2018acf}
\bibitem{Alvarado:2018acf} 
  C.~Alvarado and A.~Aranda,
  %``Standard Model Extension with Flipped Generations,''
  Phys.\ Lett.\ B {\bf 787}, 1 (2018)
%  doi:10.1016/j.physletb.2018.10.047
  [arXiv:1807.01453 [hep-ph]].

\bibitem{Hall:2018let} 
L.~J.~Hall and K.~Harigaya,
%``Implications of Higgs Discovery for the Strong CP Problem and Unification,''
JHEP {\bf 1810}, 130 (2018)
%doi:10.1007/JHEP10(2018)130
[arXiv:1803.08119 [hep-ph]].
%%CITATION = doi:10.1007/JHEP10(2018)130;%%

\bibitem{Ema:2016ops} 
Y.~Ema, K.~Hamaguchi, T.~Moroi and K.~Nakayama,
%``Flaxion: a minimal extension to solve puzzles in the standard model,''
JHEP {\bf 1701}, 096 (2017)
%doi:10.1007/JHEP01(2017)096
[arXiv:1612.05492 [hep-ph]].
%%CITATION = doi:10.1007/JHEP01(2017)096;%%

\bibitem{Ema:2018abj} 
Y.~Ema, D.~Hagihara, K.~Hamaguchi, T.~Moroi and K.~Nakayama,
%``Supersymmetric Flaxion,''
JHEP {\bf 1804}, 094 (2018)
%doi:10.1007/JHEP04(2018)094
[arXiv:1802.07739 [hep-ph]].
%%CITATION = doi:10.1007/JHEP04(2018)094;%%

\bibitem{Calibbi:2016hwq} 
L.~Calibbi, F.~Goertz, D.~Redigolo, R.~Ziegler and J.~Zupan,
%``Minimal axion model from flavor,''
Phys.\ Rev.\ D {\bf 95}, no. 9, 095009 (2017)
%doi:10.1103/PhysRevD.95.095009
[arXiv:1612.08040 [hep-ph]].

%\cite{Alanne:2018fns}
\bibitem{Alanne:2018fns} 
  T.~Alanne, S.~Blasi and F.~Goertz,
  %``A Common Source for Scalars: Axiflavon-Higgs Unification,''
  arXiv:1807.10156 [hep-ph].

%\cite{Alves:2016bib}
\bibitem{Alves:2016bib} 
  A.~Alves, D.~A.~Camargo, A.~G.~Dias, R.~Longas, C.~C.~Nishi and F.~S.~Queiroz,
  %``Collider and Dark Matter Searches in the Inert Doublet Model from Peccei-Quinn Symmetry,''
  JHEP {\bf 1610}, 015 (2016)
%  doi:10.1007/JHEP10(2016)015
  [arXiv:1606.07086 [hep-ph]].
 
 %\cite{Dev:2016xcp}
\bibitem{Dev:2016xcp} 
  P.~S.~Bhupal Dev, R.~N.~Mohapatra and Y.~Zhang,
  %``Naturally stable right-handed neutrino dark matter,''
  JHEP {\bf 1611}, 077 (2016)
%  doi:10.1007/JHEP11(2016)077
  [arXiv:1608.06266 [hep-ph]].
 
 %\cite{Perl:2001xi}
\bibitem{Perl:2001xi} 
  M.~L.~Perl, P.~C.~Kim, V.~Halyo, E.~R.~Lee, I.~T.~Lee, D.~Loomba and K.~S.~Lackner,
  %``The Search for stable, massive, elementary particles,''
  Int.\ J.\ Mod.\ Phys.\ A {\bf 16}, 2137 (2001)
%  doi:10.1142/S0217751X01003548
  [hep-ex/0102033].

 
   %\cite{DeLuca:2018mzn}
%\bibitem{DeLuca:2018mzn} 
 % V.~De Luca, A.~Mitridate, M.~Redi, J.~Smirnov and A.~Strumia,
  %``Colored Dark Matter,''
%  arXiv:1801.01135 [hep-ph].
  %%CITATION = ARXIV:1801.01135;%%
  %\cite{Chacko:2005pe}

%\cite{Kang:2006yd}
\bibitem{Kang:2006yd} 
  J.~Kang, M.~A.~Luty and S.~Nasri,
  %``The Relic abundance of long-lived heavy colored particles,''
  JHEP {\bf 0809}, 086 (2008)
%  doi:10.1088/1126-6708/2008/09/086
  [hep-ph/0611322].

%\cite{Jacoby:2007nw}
\bibitem{Jacoby:2007nw} 
  C.~Jacoby and S.~Nussinov,
  %``The Relic Abundance of Massive Colored Particles after a Late Hadronic Annihilation Stage,''
  arXiv:0712.2681 [hep-ph].

%\cite{Geller:2018biy}
\bibitem{Geller:2018biy} 
  M.~Geller, S.~Iwamoto, G.~Lee, Y.~Shadmi and O.~Telem,
  %``Dark quarkonium formation in the early universe,''
  JHEP {\bf 1806}, 135 (2018)
%  doi:10.1007/JHEP06(2018)135
  [arXiv:1802.07720 [hep-ph]].


%\cite{Aaboud:2016uth}
\bibitem{Aaboud:2016uth} 
  M.~Aaboud {\it et al.} [ATLAS Collaboration],
  %``Search for heavy long-lived charged $R$-hadrons with the ATLAS detector in 3.2 fb$^{-1}$ of proton--proton collision data at $\sqrt{s} = 13$ TeV,''
  Phys.\ Lett.\ B {\bf 760}, 647 (2016)
%  doi:10.1016/j.physletb.2016.07.042
  [arXiv:1606.05129 [hep-ex]].
  %%CITATION = doi:10.1016/j.physletb.2016.07.042;%%
  %43 citations counted in INSPIRE as of 04 Dec 2018


%\cite{Dolgov:2013una}
\bibitem{Dolgov:2013una} 
  A.~D.~Dolgov, S.~L.~Dubovsky, G.~I.~Rubtsov and I.~I.~Tkachev,
  %``Constraints on millicharged particles from Planck data,''
  Phys.\ Rev.\ D {\bf 88}, no. 11, 117701 (2013)
%  doi:10.1103/PhysRevD.88.117701
  [arXiv:1310.2376 [hep-ph]].
  
%\cite{Abe:2016wck}
\bibitem{Abe:2016wck} 
  T.~Abe, J.~Kawamura, S.~Okawa and Y.~Omura,
  %``Dark matter physics, flavor physics and LHC constraints in the dark matter model with a bottom partner,''
  JHEP {\bf 1703}, 058 (2017)
%  doi:10.1007/JHEP03(2017)058
  [arXiv:1612.01643 [hep-ph]].
  %%CITATION = doi:10.1007/JHEP03(2017)058;%%
  %6 citations counted in INSPIRE as of 13 Dec 2018
  
%\cite{Okawa:2017bgh}
\bibitem{Okawa:2017bgh} 
  S.~Okawa and Y.~Omura,
  %``Hidden sector behind the CKM matrix,''
  Phys.\ Rev.\ D {\bf 96}, no. 3, 035012 (2017)
 % doi:10.1103/PhysRevD.96.035012
  [arXiv:1703.08789 [hep-ph]].
  %%CITATION = doi:10.1103/PhysRevD.96.035012;%%
  %3 citations counted in INSPIRE as of 13 Dec 2018


\bibitem{Chacko:2005pe} 
  Z.~Chacko, H.~S.~Goh and R.~Harnik,
  %``The Twin Higgs: Natural electroweak breaking from mirror symmetry,''
  Phys.\ Rev.\ Lett.\  {\bf 96}, 231802 (2006)
%  doi:10.1103/PhysRevLett.96.231802
  [hep-ph/0506256].
  %%CITATION = doi:10.1103/PhysRevLett.96.231802;%%
  %394 citations counted in INSPIRE as of 27 Nov 2018
  
  %\cite{Ellis:1978hq}
\bibitem{Ellis:1978hq} 
J.~R.~Ellis and M.~K.~Gaillard,
%``Strong and Weak CP Violation,''
Nucl.\ Phys.\ B {\bf 150}, 141 (1979).
%doi:10.1016/0550-3213(79)90297-9
%%CITATION = doi:10.1016/0550-3213(79)90297-9;%%
%219 citations counted in INSPIRE as of 17 May 2018

%%%%%%%%%% Higgs portal DM %%%%%%%%%%%
%\cite{Kanemura:2010sh}
\bibitem{Kanemura:2010sh} 
  S.~Kanemura, S.~Matsumoto, T.~Nabeshima and N.~Okada,
  %``Can WIMP Dark Matter overcome the Nightmare Scenario?,''
  Phys.\ Rev.\ D {\bf 82}, 055026 (2010)
%  doi:10.1103/PhysRevD.82.055026
  [arXiv:1005.5651 [hep-ph]].

%\cite{Djouadi:2011aa}
\bibitem{Djouadi:2011aa} 
  A.~Djouadi, O.~Lebedev, Y.~Mambrini and J.~Quevillon,
  %``Implications of LHC searches for Higgs--portal dark matter,''
  Phys.\ Lett.\ B {\bf 709}, 65 (2012)
%  doi:10.1016/j.physletb.2012.01.062
  [arXiv:1112.3299 [hep-ph]].

%\cite{Djouadi:2012zc}
\bibitem{Djouadi:2012zc} 
  A.~Djouadi, A.~Falkowski, Y.~Mambrini and J.~Quevillon,
  %``Direct Detection of Higgs-Portal Dark Matter at the LHC,''
  Eur.\ Phys.\ J.\ C {\bf 73}, no. 6, 2455 (2013)
%  doi:10.1140/epjc/s10052-013-2455-1
  [arXiv:1205.3169 [hep-ph]].

%\cite{Escudero:2016gzx}
\bibitem{Escudero:2016gzx} 
  M.~Escudero, A.~Berlin, D.~Hooper and M.~X.~Lin,
  %``Toward (Finally!) Ruling Out Z and Higgs Mediated Dark Matter Models,''
  JCAP {\bf 1612}, 029 (2016)
%  doi:10.1088/1475-7516/2016/12/029
  [arXiv:1609.09079 [hep-ph]].

%\cite{Balazs:2017ple}
\bibitem{Balazs:2017ple} 
  J.~Ellis, A.~Fowlie, L.~Marzola and M.~Raidal,
  %``Statistical Analyses of Higgs- and Z-Portal Dark Matter Models,''
  Phys.\ Rev.\ D {\bf 97}, no. 11, 115014 (2018)
%  doi:10.1103/PhysRevD.97.115014
  [arXiv:1711.09912 [hep-ph]].

%\cite{Athron:2018ipf}
\bibitem{Athron:2018ipf} 
  P.~Athron, J.~M.~Cornell, F.~Kahlhoefer, J.~Mckay, P.~Scott and S.~Wild,
  %``Impact of vacuum stability, perturbativity and XENON1T on global fits of $\mathbb {Z}_2$ and $\mathbb {Z}_3$ scalar singlet dark matter,''
  Eur.\ Phys.\ J.\ C {\bf 78}, no. 10, 830 (2018)
%  doi:10.1140/epjc/s10052-018-6314-y
  [arXiv:1806.11281 [hep-ph]].
%%%%%%%%%%%%%%%%%%%%%%


%\cite{Hisano:2015bma}
\bibitem{Hisano:2015bma} 
  J.~Hisano, R.~Nagai and N.~Nagata,
  %``Effective Theories for Dark Matter Nucleon Scattering,''
  JHEP {\bf 1505}, 037 (2015)
%  doi:10.1007/JHEP05(2015)037
  [arXiv:1502.02244 [hep-ph]].

%\cite{Aprile:2018dbl}
\bibitem{Aprile:2018dbl} 
  E.~Aprile {\it et al.} [XENON Collaboration],
  %``Dark Matter Search Results from a One Ton-Year Exposure of XENON1T,''
  Phys.\ Rev.\ Lett.\  {\bf 121}, no. 11, 111302 (2018)
%  doi:10.1103/PhysRevLett.121.111302
  [arXiv:1805.12562 [astro-ph.CO]].

%\cite{Akerib:2016vxi}
\bibitem{Akerib:2016vxi} 
  D.~S.~Akerib {\it et al.} [LUX Collaboration],
  %``Results from a search for dark matter in the complete LUX exposure,''
  Phys.\ Rev.\ Lett.\  {\bf 118}, no. 2, 021303 (2017)
%  doi:10.1103/PhysRevLett.118.021303
  [arXiv:1608.07648 [astro-ph.CO]].

%%%%%%%%%% top partner %%%%%%%%%%%
%\cite{Baek:2016lnv}
\bibitem{Baek:2016lnv} 
  S.~Baek, P.~Ko and P.~Wu,
  %``Top-philic Scalar Dark Matter with a Vector-like Fermionic Top Partner,''
  JHEP {\bf 1610}, 117 (2016)
%  doi:10.1007/JHEP10(2016)117
  [arXiv:1606.00072 [hep-ph]].

%\cite{Blanke:2017tnb}
\bibitem{Blanke:2017tnb} 
  M.~Blanke and S.~Kast,
  %``Top-Flavoured Dark Matter in Dark Minimal Flavour Violation,''
  JHEP {\bf 1705}, 162 (2017)
%  doi:10.1007/JHEP05(2017)162
  [arXiv:1702.08457 [hep-ph]].
%%%%%%%%%%%%%%%%%%%%%% 

  %%%%%%%%%input%%%%%%%%%%%%%%%%

\bibitem{PDG}
  C.~Patrignani {\it et al.} [Particle Data Group],
  %``Review of Particle Physics,''
  Chin.\ Phys.\ C {\bf 40}, no. 10, 100001 (2016).
%  doi:10.1088/1674-1137/40/10/100001
  %%CITATION = doi:10.1088/1674-1137/40/10/100001;%%
  %403 citations counted in INSPIRE as of 24 Feb 2017

  %\cite{Amhis:2016xyh}
\bibitem{HFLAV} 
  Y.~Amhis {\it et al.} [HFLAV Collaboration],
  %``Averages of $b$-hadron, $c$-hadron, and $\tau$-lepton properties as of summer 2016,''
  Eur.\ Phys.\ J.\ C {\bf 77}, no. 12, 895 (2017)
  %doi:10.1140/epjc/s10052-017-5058-4
  [arXiv:1612.07233 [hep-ex]].
  %%CITATION = doi:10.1140/epjc/s10052-017-5058-4;%%
  %306 citations counted in INSPIRE as of 16 May 2018

%\cite{Falkowski:2013jya}
\bibitem{Falkowski:2013jya} 
  A.~Falkowski, D.~M.~Straub and A.~Vicente,
  %``Vector-like leptons: Higgs decays and collider phenomenology,''
  JHEP {\bf 1405}, 092 (2014)
%  doi:10.1007/JHEP05(2014)092
  [arXiv:1312.5329 [hep-ph]].
  %%CITATION = doi:10.1007/JHEP05(2014)092;%%
  %103 citations counted in INSPIRE as of 01 Dec 2018

%\cite{CMS:2018cgi}
\bibitem{CMS:2018cgi} 
  CMS Collaboration [CMS Collaboration],
  %``Search for vector-like leptons in multilepton final states in pp collisions at $\sqrt{s} = 13~\mathrm{TeV}$,''
  CMS-PAS-EXO-18-005.
  %%CITATION = CMS-PAS-EXO-18-005;%%
  %1 citations counted in INSPIRE as of 02 Dec 2018

%\cite{Ellis:2014dza}
\bibitem{Ellis:2014dza} 
  S.~A.~R.~Ellis, R.~M.~Godbole, S.~Gopalakrishna and J.~D.~Wells,
  %``Survey of vector-like fermion extensions of the Standard Model and their phenomenological implications,''
  JHEP {\bf 1409}, 130 (2014)
%  doi:10.1007/JHEP09(2014)130
  [arXiv:1404.4398 [hep-ph]].
  %%CITATION = doi:10.1007/JHEP09(2014)130;%%
  %88 citations counted in INSPIRE as of 01 Dec 2018

%\cite{Dermisek:2014qca}
\bibitem{Dermisek:2014qca} 
  R.~Dermisek, J.~P.~Hall, E.~Lunghi and S.~Shin,
  %``Limits on Vectorlike Leptons from Searches for Anomalous Production of Multi-Lepton Events,''
  JHEP {\bf 1412}, 013 (2014)
%  doi:10.1007/JHEP12(2014)013
  [arXiv:1408.3123 [hep-ph]].
  %%CITATION = doi:10.1007/JHEP12(2014)013;%%
  %35 citations counted in INSPIRE as of 01 Dec 2018

%\cite{Kumar:2015tna}
\bibitem{Kumar:2015tna} 
  N.~Kumar and S.~P.~Martin,
  %``Vectorlike Leptons at the Large Hadron Collider,''
  Phys.\ Rev.\ D {\bf 92}, no. 11, 115018 (2015)
 % doi:10.1103/PhysRevD.92.115018
  [arXiv:1510.03456 [hep-ph]].
  %%CITATION = doi:10.1103/PhysRevD.92.115018;%%
  %37 citations counted in INSPIRE as of 01 Dec 2018

%\cite{Aaboud:2018zjf}
\bibitem{Aaboud:2018zjf} 
  M.~Aaboud {\it et al.} [ATLAS Collaboration],
  %``Search for supersymmetry in final states with charm jets and missing transverse momentum in 13 TeV $pp$ collisions with the ATLAS detector,''
  JHEP {\bf 1809}, 050 (2018)
%  doi:10.1007/JHEP09(2018)050
  [arXiv:1805.01649 [hep-ex]].
  %%CITATION = doi:10.1007/JHEP09(2018)050;%%
  %7 citations counted in INSPIRE as of 03 Dec 2018  

%\cite{Sirunyan:2017kiw}
\bibitem{Sirunyan:2017kiw} 
  A.~M.~Sirunyan {\it et al.} [CMS Collaboration],
  %``Search for the pair production of third-generation squarks with two-body decays to a bottom or charm quark and a neutralino in proton\UTF{2013}proton collisions at $\sqrt{s}$ = 13 TeV,''
  Phys.\ Lett.\ B {\bf 778}, 263 (2018)
%  doi:10.1016/j.physletb.2018.01.012
  [arXiv:1707.07274 [hep-ex]].
  %%CITATION = doi:10.1016/j.physletb.2018.01.012;%%
  %24 citations counted in INSPIRE as of 03 Dec 2018

%
\bibitem{Borschensky:2014cia} 
  C.~Borschensky, M.~Kr\"amer, A.~Kulesza, M.~Mangano, S.~Padhi, T.~Plehn and X.~Portell,
  %``Squark and gluino production cross sections in pp collisions at $\sqrt{s}$ = 13, 14, 33 and 100 TeV,''
  Eur.\ Phys.\ J.\ C {\bf 74}, no. 12, 3174 (2014)
  %doi:10.1140/epjc/s10052-014-3174-y
  [arXiv:1407.5066 [hep-ph]].
  %%CITATION = doi:10.1140/epjc/s10052-014-3174-y;%%
  %257 citations counted in INSPIRE as of 03 Dec 2018


%
\bibitem{Aaboud:2017vwy} 
  M.~Aaboud {\it et al.} [ATLAS Collaboration],
  %``Search for squarks and gluinos in final states with jets and missing transverse momentum using 36\UTF{2009}\UTF{2009}fb$^{-1}$ of $\sqrt{s}=13$\UTF{2009}\UTF{2009}TeV pp collision data with the ATLAS detector,''
  Phys.\ Rev.\ D {\bf 97}, no. 11, 112001 (2018)
%  doi:10.1103/PhysRevD.97.112001
  [arXiv:1712.02332 [hep-ex]].
  %%CITATION = doi:10.1103/PhysRevD.97.112001;%%
  %56 citations counted in INSPIRE as of 03 Dec 2018
%\cite{Sirunyan:2018vjp}
\bibitem{Sirunyan:2018vjp} 
  A.~M.~Sirunyan {\it et al.} [CMS Collaboration],
  %``Search for natural and split supersymmetry in proton-proton collisions at $ \sqrt{s}=13 $ TeV in final states with jets and missing transverse momentum,''
  JHEP {\bf 1805}, 025 (2018)
%  doi:10.1007/JHEP05(2018)025
  [arXiv:1802.02110 [hep-ex]].
  %%CITATION = doi:10.1007/JHEP05(2018)025;%%
  %22 citations counted in INSPIRE as of 03 Dec 2018

\bibitem{Aaboud:2018hdl} 
  M.~Aaboud {\it et al.} [ATLAS Collaboration],
  %``Search for heavy charged long-lived particles in proton-proton collisions at $\sqrt{s} = 13$ TeV using an ionisation measurement with the \mbox{ATLAS} detector,''
%  doi:10.1016/j.physletb.2018.10.055
  arXiv:1808.04095 [hep-ex].
  %%CITATION = doi:10.1016/j.physletb.2018.10.055;%%
  %2 citations counted in INSPIRE as of 24 Nov 2018

%\cite{TheMEG:2016wtm}
\bibitem{TheMEG:2016wtm} 
  A.~M.~Baldini {\it et al.} [MEG Collaboration],
  %``Search for the lepton flavour violating decay $\mu ^+ \rightarrow \mathrm {e}^+ \gamma $ with the full dataset of the MEG experiment,''
  Eur.\ Phys.\ J.\ C {\bf 76}, no. 8, 434 (2016)
%  doi:10.1140/epjc/s10052-016-4271-x
  [arXiv:1605.05081 [hep-ex]].
  %%CITATION = doi:10.1140/epjc/s10052-016-4271-x;%%
  %184 citations counted in INSPIRE as of 16 May 2018


%%%%%%%%%%%%%%muto3e%%%%%%%%%%%%%
\bibitem{Bellgardt:1987du}
  U.~Bellgardt {\it et al.}  [SINDRUM Collaboration],
  %``Search for the Decay $\mu^+ \to e^+ e^+ e^-$,''
  Nucl.\ Phys.\ B {\bf 299}, 1 (1988).
  %%CITATION = NUPHA,B299,1;%%
  %370 citations counted in INSPIRE as of 10 Nov 2014

%\cite{Perrevoort:2016nuv}
\bibitem{Perrevoort:2016nuv} 
  A.~K.~Perrevoort [Mu3e Collaboration],
  %``Status of the Mu3e Experiment at PSI,''
  EPJ Web Conf.\  {\bf 118}, 01028 (2016)
 % doi:10.1051/epjconf/201611801028
  [arXiv:1605.02906 [physics.ins-det]].
  %%CITATION = doi:10.1051/epjconf/201611801028;%%
  %10 citations counted in INSPIRE as of 07 Mar 2018

%%%%%%%%%%%%%%%%%%%%%%


%\cite{Aaboud:2017leg}
\bibitem{Aaboud:2017leg} 
  M.~Aaboud {\it et al.} [ATLAS Collaboration],
  %``Search for electroweak production of supersymmetric states in scenarios with compressed mass spectra at $\sqrt{s}=13$ TeV with the ATLAS detector,''
  Phys.\ Rev.\ D {\bf 97}, no. 5, 052010 (2018)
%  doi:10.1103/PhysRevD.97.052010
  [arXiv:1712.08119 [hep-ex]].
  %%CITATION = doi:10.1103/PhysRevD.97.052010;%%
  %44 citations counted in INSPIRE as of 03 Dec 2018

%\cite{Aaboud:2018jiw}
\bibitem{Aaboud:2018jiw} 
  M.~Aaboud {\it et al.} [ATLAS Collaboration],
  %``Search for electroweak production of supersymmetric particles in final states with two or three leptons at $\sqrt{s}=13\,$TeV with the ATLAS detector,''
  arXiv:1803.02762 [hep-ex].
  %%CITATION = ARXIV:1803.02762;%%
  %48 citations counted in INSPIRE as of 03 Dec 2018
%\cite{Sirunyan:2018nwe}
\bibitem{Sirunyan:2018nwe} 
  A.~M.~Sirunyan {\it et al.} [CMS Collaboration],
  %``Search for supersymmetric partners of electrons and muons in proton-proton collisions at $\sqrt{s}=$ 13 TeV,''
  %Submitted to: Phys.Lett.
  [arXiv:1806.05264 [hep-ex]].
  %%CITATION = ARXIV:1806.05264;%%
  %8 citations counted in INSPIRE as of 03 Dec 2018

%\cite{Sirunyan:2018vig}
\bibitem{Sirunyan:2018vig} 
  A.~M.~Sirunyan {\it et al.} [CMS Collaboration],
  %``Search for supersymmetry in events with a $\tau$ lepton pair and missing transverse momentum in proton-proton collisions at $\sqrt{s} =$ 13 TeV,''
  JHEP {\bf 1811}, 151 (2018)
%  doi:10.1007/JHEP11(2018)151
  [arXiv:1807.02048 [hep-ex]].
  %%CITATION = doi:10.1007/JHEP11(2018)151;%%
  %8 citations counted in INSPIRE as of 03 Dec 2018

%\cite{Aaboud:2018pii}
\bibitem{Aaboud:2018pii} 
  M.~Aaboud {\it et al.} [ATLAS Collaboration],
  %``Combination of the searches for pair-produced vector-like partners of the third-generation quarks at $\sqrt{s} =$ 13 TeV with the ATLAS detector,''
  Phys.\ Rev.\ Lett.\  {\bf 121}, no. 21, 211801 (2018)
%  doi:10.1103/PhysRevLett.121.211801
  [arXiv:1808.02343 [hep-ex]].
  %%CITATION = doi:10.1103/PhysRevLett.121.211801;%%
  %9 citations counted in INSPIRE as of 03 Dec 2018
%\cite{CMS:2018haz}
\bibitem{CMS:2018haz} 
  CMS Collaboration [CMS Collaboration],
  %``Search for a vector-like quark decaying to a top quark and a W boson,''
  CMS-PAS-B2G-17-018.
  %%CITATION = CMS-PAS-B2G-17-018;%%
%\cite{Sirunyan:2018ncp}
%\bibitem{Sirunyan:2018ncp} 
%  A.~M.~Sirunyan {\it et al.} [CMS Collaboration],
  %``Search for single production of vector-like quarks decaying to a top quark and a W boson in proton-proton collisions at $\sqrt{s} =$ 13 TeV,''
%  arXiv:1809.08597 [hep-ex].
  %%CITATION = ARXIV:1809.08597;%%
  %3 citations counted in INSPIRE as of 03 Dec 2018

\bibitem{Planck:2018eyx} 
N.~Aghanim {\it et al.} [Planck Collaboration],
%``Planck 2018 results. VI. Cosmological parameters,''
arXiv:1807.06209 [astro-ph.CO].
%%CITATION = ARXIV:1807.06209;%%
%294 citations counted in INSPIRE as of 28 Nov 2018

\bibitem{Abazajian:2016yjj} 
K.~N.~Abazajian {\it et al.} [CMB-S4 Collaboration],
%``CMB-S4 Science Book, First Edition,''
arXiv:1610.02743 [astro-ph.CO].
%%CITATION = ARXIV:1610.02743;%%
%338 citations counted in INSPIRE as of 25 Sep 2018
  


\end{thebibliography}
\end{document}